\documentclass[12pt]{iopart}

\expandafter\let\csname equation*\endcsname\relax
\expandafter\let\csname endequation*\endcsname\relax 
\usepackage{amsmath}
\usepackage{amssymb}
\usepackage{epsfig}
\usepackage{color}
\usepackage{bbm}

\graphicspath{{.}{./EPS/}}

\newcommand* {\bra}[1]{\ensuremath{\langle {#1} |}}
\newcommand* {\ket}[1]{\ensuremath{| {#1} \rangle}}
\newcommand* {\ee}{\ensuremath{\mathrm{e}}}

\begin{document}
 
\title[]{Centre-of-mass motion-induced decoherence and entanglement generation
 in a hybrid quantum repeater}
\author{J Z Bern\'ad, H Frydrych and G Alber}
\address{Institut f\"{u}r Angewandte Physik, Technische Universit\"{a}t Darmstadt, D-64289, Germany}
\ead{Zsolt.Bernad@physik.tu-darmstadt.de}

\date{\today}

\begin{abstract}

Quantum communication over long distances relies on the ability to create entanglement between two remote quantum nodes. 
Recent proposals aiming at experimental realization propose a hybrid quantum repeater setup where two distant material qubits are entangled 
by light-matter interaction. Motivated by these developments, we investigate possible decoherence effects originating from the centre-of-mass 
motion of the spatially well separated trapped qubits.
Within the Lamb-Dicke regime we use photon exchange involving coherent states of the radiation field to entangle the two material qubits. 
Optimal generalized photonic field measurements are used to achieve entangled qubit pairs with high fidelities and high success 
probabilities.
We demonstrate that the quality of the achievable
two-qubit entanglement crucially depends on the trap frequencies involved.  
Furthermore, 
dynamical decoupling schemes are proposed which 
are capable of
suppressing 
centre-of-mass-motion-induced decoherence effects significantly
and which 
involve only local operations acting on the spatially well-separated material qubits.
\end{abstract}

\submitto{\jpb}

\maketitle

\section{Introduction}
Reliable entanglement distribution between quantum nodes
over long distances is of crucial importance for quantum communication.
A possible way of overcoming the destructive influence of
decoherence in the process of entanglement distribution
is provided by  quantum repeaters \cite{Briegel98,Duer99}. They
take advantage of previously shared entanglement between 
neighbouring pairs of quantum nodes and enable the generation of entanglement between two
distant quantum nodes by
entanglement swapping \cite{Zuk}. 
Furthermore, subsequent entanglement purification procedures \cite{Bennett,Deutsch} 
are capable of distilling high-fidelity entangled pairs from 
a sufficiently
large number of low-fidelity entangled pairs. Recently various physical setups and
entanglement distribution protocols have been proposed
for the realization of quantum repeaters \cite{Sangouard}.


Implementations of entanglement distribution which are
compatible with existing classical optical communication networks and which are based on
multiphoton signals are particularly attractive. 
The recent proposal of van Loock et al. \cite{vanLoock1,vanLoock2,vanLoock3} of a hybrid
quantum repeater is an example for such an implementation. It takes advantage of
the transmission of coherent 
photon states through an optical fibre and subsequent photonic postselection
for the generation of entanglement between distant pairs of material qubits which are
entangled with the photons by weak non-resonant interactions.
However, in this proposal
the assumed weak non-resonant couplings between the material qubits and the photons 
involved impose unfavourable restrictions on
the photonic postselection process by which the material entangled qubit pairs are prepared. 
In order to overcome these limitations,
recently
this hybrid quantum repeater model has been generalized to the resonant
strong coupling regime \cite{Bernad}.  
It has been demonstrated that
the collapse phenomenon well known from the Jaynes-Cummings-Paul model \cite{JCP} 
can cause favourable quantum correlations between two material qubits and the photons involved,
which enable the preparation of
perfectly entangled material qubit pairs by photonic postselection with the help of a
von Neumann measurement. In this idealized model it has been assumed that
effects of spontaneous emission during the short qubit-photon interactions 
are negligible and that
photonic states can be transferred perfectly between the two cavities
containing the two trapped material qubits by an interconnecting long-distance optical fibre.
Although a realization of the assumed cavity-fibre couplings is still challenging,
recently highly promising experimental developments
have been taking place in this direction \cite{Steinmetz,Colombe}.
Furthermore, effects originating from the motion of the trapped material qubits 
on the entanglement generation have been neglected in these early investigations
in the strong-coupling limit.

An important issue in any implementation of such a photon-mediated
entanglement distribution
scenario
is the physical realization of the material qubits \cite{Sangouard2}. Trapped ions or atoms are well suited
for this purpose as the quantum technology for controlling their degrees of freedom 
is already well advanced
\cite{Leibfried,L2,Blatt}. In particular, it is important to control the centre-of-mass motion
of these material qubits properly as it introduces unwanted decoherence and dissipation.
Motivated by the current interest in realizations of hybrid quantum repeaters and
in entangelement distribution in general,
in this paper the decohering influence of the centre of motion of trapped qubits on 
entanglement generation is investigated.
For this purpose we start from our previous generalization of the quantum repeater model
of van Loock et al. \cite{vanLoock1} and discuss the decohering and dissipative
influence of the qubits' centre-of-mass
motion on the generation of distant qubit entanglement in the strong quantum electrodynamical
coupling regime. 
We demonstrate that the quality of the generated entanglement is 
rather sensitive to the trap frequency, with
high trap frequencies increasing the fidelity of the entangled pairs.
A second main aim of our work is to propose a local dynamical decoupling scheme 
\cite{Viola1,Viola2,Zanardi,Vitali} which
is capable of eliminating the unwanted influence of the centre-of-mass motion 
by acting on each
trapped material qubit only locally.
This dynamical decoupling scheme generalizes previous ideas of
Facchi et al. \cite{facchi} to unitary decoupling operations with degenerate spectra acting on an
infinite dimensional Hilbert space.
Furthermore, the fact that
this dynamical decoupling scheme does not act on the internal degrees of freedom of the locally trapped qubits directly,
but only on their centre-of-mass degrees of freedoms makes it 
attractive for potential experimental realizations.

The outline of the paper is as follows.
In section \ref{The model} we introduce our theoretical model. The internal
physical degrees of freedom of the two trapped
qubits to be entangled are modelled by 
three-level systems whose
upper electronic levels are coupled
by single-mode photon fields inside cavities. The two photon cavities containing these
trapped material three-level systems are  coupled by a long optical fibre.
Furthermore, we shortly describe the general 
framework of optimal generalized photon measurements, which lead to entanglement of the
two material quantum systems by photonic postselection. Numerical results are presented 
describing the decohering and dissipating influence of the trapped qubits' centre-of-mass motion
on the fidelity of the generated entangled qubit states and on the relevant
success and minimum error probabilities. 
In section \ref{decoupling} we address the suppression of these decohering and dissipating effects
and propose
a dynamical decoupling scheme which acts only on the degrees of freedom of the centre-of-mass motion.
Technical details, such as the derivation of
a proper Baker-Hausdorff formula and details of the relevant photon states, are given in \ref{BH} and \ref{Calc}.

\section{Entanglement generation in the presence of centre-of-mass motion}
\label{The model}
In a hybrid quantum repeater, entanglement is created between two distant material qubits
with the help of photon exchange and subsequent photonic postselection.
The two spatially separated material qubits can, for example, be implemented as internal states of trapped ions or atoms. 
In a Ramsey-type interaction scenario, the first qubit interacts shortly
with the radiation field inside a cavity 
resulting in an entangled state between this qubit and the photon field. After transmission of the
resulting photon wave packet into a second distant cavity by an optical fibre
it interacts shortly with the second qubit.
If the photon state transfer between both cavities is perfect,
the resulting entanglement between
the two distant material qubits and the 
photons can be used to
prepare an almost perfectly entangled state between the two qubits by an appropriate photonic measurement. 
A recent theoretical investigation
demonstrates that such perfect photonic quantum state transfer between two distant
cavities is possible by an appropriate choice of the couplings between two cavities by an optical fibre
\cite{Bernad} (see also \cite{Cirac,Enk,Pellizzari} for alternative solutions) and recent experimental developments 
\cite{Steinmetz,Colombe} indicate that such photonic quantum state transfers can be realized.  
In general, however, in such a scenario the centre-of-mass motion has to be taken into
account as its degrees of freedom also participate in the formation of the entanglement between
the material quantum systems and the photons involved, thus causing decoherence and dissipation.
The main aim of this chapter is to explore this particular source of
decoherence and dissipation which has been neglected in investigations so far.
In particular we are interested in the circumstances under which
high-fidelity entangled states of the distant qubits can be prepared by optimal photon measurements. 
\begin{figure}[b]
\begin{center}
\includegraphics[width=7cm]{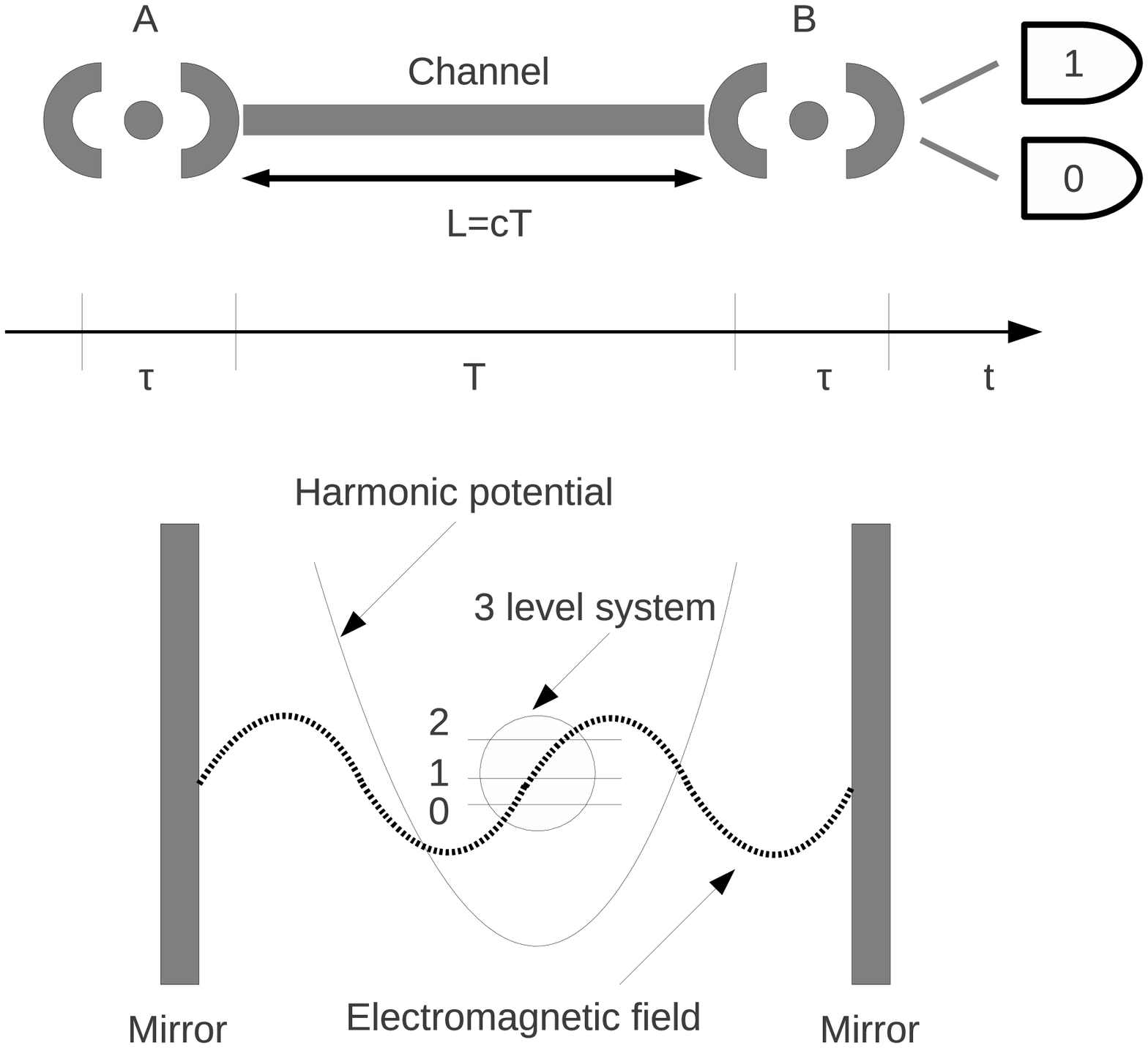}
\includegraphics[width=4.5cm]{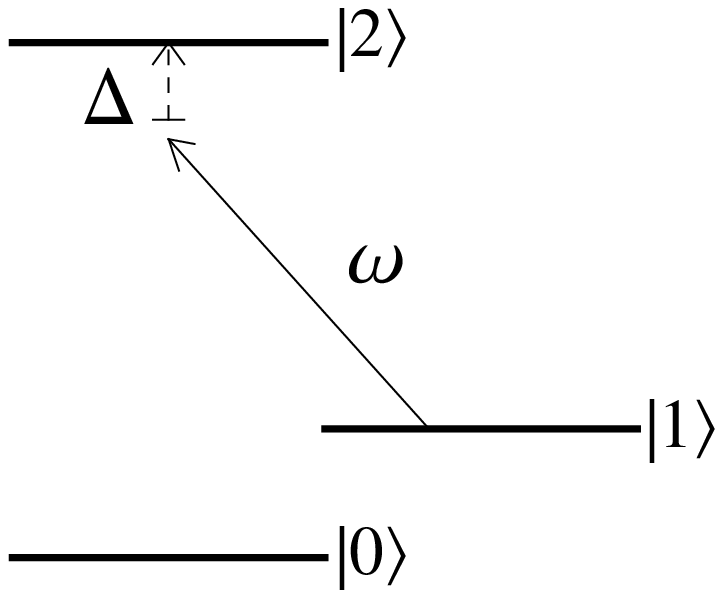}
\caption{\label{fig-1} 
Generation of photon-assisted entanglement:
The first interaction of duration $\tau$  results in an entangled state between the material quantum system and the radiation field
in cavity A; after transferring the photons by an optical fibre with length $L$ into the second cavity $B$, the propagated photons interact 
in a similar way with the second material quantum system $B$. Finally the resulted state of the radiation field is projected by a 
minimum-error two-valued POVM measurement with measurement results $1$ or $0$. The measurement result $1$ prepares both material quantum 
systems approximately in a Bell state $|\Psi^+\rangle$ of states $\ket{0}$ and $\ket{1}$ with success probability $P_{Bell}$ and with 
fidelity $F_{opt}$. Schematic diagrams of the structure of internal states and their coupling to the centre-of-mass motion are depicted.
}
\end{center}  
\end{figure}  
\subsection{Model Hamiltonian for the interaction}
\label{modelH}

We consider a three-level trapped system (ion or atom) in a harmonic potential with frequency $\omega_t$, placed inside an optical 
cavity with frequency 
$\omega_c$.
The system is considered to be located at the origin. The internal energy eigenstates are  $\ket{0}$, $\ket{1}$ and $\ket{2}$ with associated energies $E_0$, $E_1$ and
$E_2$. The internal states are treated as a ladder system with two hyperfine-split components 
$\ket{0}$ and $\ket{1}$ acting as the qubit states, of which only the state $\ket{1}$ participates in the interaction with the cavity mode 
and the centre-of-mass motion. 
These two states have long radiative lifetimes. 

Assuming that the electric field does not change considerably over the size of 
the atom or ion, the total Hamiltonian in the dipole approximation reads \cite{JCP}
\begin{eqnarray}
 \hat{H}&=&\frac{\hat{p}^2}{2m}+\frac{m \omega^2_t \hat{x}^2}{2}+\hbar \omega_0 \ket{0} \bra{0}+ \frac{1}{2}\hbar \omega_{21} \hat{\sigma}_z+
\hbar \omega_c \hat{a}^\dagger \hat{a} \nonumber \\
&+&\hbar \big(\hat{\sigma}_+ + \hat{\sigma}_-\big) \big(g^*(k\hat{x})\hat{a}^\dagger+g(k\hat{x})\hat{a}\big),
\end{eqnarray}
where $\hat{\sigma}_z=\ket{2} \bra{2}-\ket{1} \bra{1}$,$\hat{\sigma}_+=\ket{2} \bra{1}$ and $\hat{\sigma}_-=\ket{1} \bra{2}$ with 
$\hbar \omega_{21}=(E_2-E_1)/2$ and $\hbar \omega_0=E_0+(E_1+E_2)/2$. The Hamiltonian includes the kinetic energy operator 
$\frac{\hat{p}^2}{2m}$ of the
centre-of-mass motion with mass $m$ in the harmonic potential $\frac{m \omega^2_t \hat{x}^2}{2}$. 
$\hat{a}$ ($\hat{a}^\dagger$) is the destruction (creation) operator of the electromagnetic field mode. The coupling
operator $g(k\hat{x})$ characterizes the strength of the interaction of the material system with the single-mode 
of the radiation field and is given by
\begin{equation}
\hbar g(k\hat{x})=-\sqrt{\frac{\hbar \omega_c}{2 \epsilon_0}}\bra{1} \hat{\vec{d}} 
\ket{2}\cdot \vec{u}(\hat{x},0,0),
\end{equation}
where $\hat{\vec{d}}$ is the dipole operator and $k$ is the wave number of the field. ($\epsilon_0$ is the permittivity of vacuum.) 
The normalized mode function $\vec{u}(\vec{r})$ is a solution to the Helmholtz
equation 
\begin{equation}
 \big(\nabla^2+\frac{\omega^2_c}{c^2} \big)\vec{u}(\vec{r})=0
\end{equation}
and fulfills the boundary conditions of the cavity and the Coulomb gauge condition.

We now define the position and momentum operator in terms of the annihilation and creation operators $\hat{b}$ and $\hat{b}^\dagger$, 
that is
\begin{eqnarray}
 \hat{x}=\sqrt{\frac{\hbar}{2 m \omega_t}} \big(\hat{b}+\hat{b}^\dagger \big), \,\,\,\,\,\,\,
 \hat{p}=\sqrt{\frac{m \hbar \omega_t}{2}} \frac{1}{i}\big(\hat{b}-\hat{b}^\dagger \big).
\end{eqnarray}
The minimum of the harmonic potential is in the position $x=0$ and we are going to Taylor expand the coupling operator around 
this point. There are two necessary conditions to justify this expansion, namely the function $g(x)$ is smooth in the neighbourhood of 
the origin and the Lamb-Dicke parameter
\begin{equation}
 \eta=k\sqrt{\frac{\hbar}{2 m \omega_t}} \ll 1
\label{eta}
\end{equation}
is small. The smoothness of $g(x)$ is guaranteed by $\vec{u}$, which is a solution to the Helmholtz equation. 
The Lamb-Dicke parameter $\eta$ 
measures the deviation
\begin{equation}
 \Delta x = \sqrt{\frac{\hbar}{2m \omega_t}}
\end{equation}
of the centre-of-mass motion with respect to the wave length $\lambda$ of the field 
\begin{equation}
 \eta=k \sqrt{\frac{\hbar}{2m \omega_t}} = 2 \pi \frac{\Delta x}{\lambda}.
\end{equation}
As an example, in Ref. \cite{Steiner} the wavelength of a single-mode cavity is around $\lambda=935$ $nm$, and 
inside this cavity a $\mbox{Yb}^{+}$ ion is trapped in an approximately $1.5-2.5$ MHz oscillatory potential, which 
yields a Lamb-Dicke parameter $\eta \sim 10^{-1}$. 

Since both conditions for the Taylor expansion are fulfilled, the coupling operator can be written as
\begin{equation}
 g(k \hat{x})\cong g + \eta g'(x)\mid_{x=0} \big(\hat{b}+\hat{b}^\dagger\big).
\end{equation}
With the help of the rotating wave approximation for the interaction between the radiation field and the internal states we arrive at
\begin{eqnarray}
  \hat{H}&=&\hbar \omega_t \hat{b}^\dagger \hat{b}+\hbar \omega_0 \ket{0} \bra{0}+ \frac{1}{2}\hbar \omega_{21} \hat{\sigma}_z+
\hbar g \hat{\sigma}_+ \hat{a} + \hbar g^* \hat{\sigma}_- \hat{a}^\dagger \nonumber \\
&+&\hbar \omega_c \hat{a}^\dagger \hat{a}+ 
\hbar \gamma \hat{\sigma}_+ \hat{a} \big(\hat{b}+\hat{b}^\dagger\big) + \hbar \gamma^* \hat{\sigma}_- \hat{a}^\dagger 
\big(\hat{b}+\hat{b}^\dagger\big),
\end{eqnarray}
where $\gamma=\eta g'(0)$.

We describe this dynamical system by 
introducing dressed states as eigenstates of the internal states $\ket{1}$, $\ket{2}$ and the radiation field Hamiltonian.
In order to use these dressed states effectively, we have to transform our system into an interaction picture where the interaction 
Hamiltonian between the field and the internal states is time independent. To do so, we apply the unitary transformation
\begin{eqnarray}
\hat{U}(t)&=&\ee^{i/\hbar \hat{H}_S t}, \nonumber \\
\hat{H}_S&=&\hbar \omega_0 \ket{0} \bra{0}+\frac{1}{2} \hbar \omega_c \sigma_z+ \hbar \omega_c \hat{a}^\dagger \hat{a}.
\end{eqnarray}
The Hamiltonian in this interaction picture reads
\begin{eqnarray}
\label{modelI1}
\hat{H}_I&=& \hbar \omega_t \hat{b}^\dagger \hat{b}+\frac{1}{2}\hbar \Delta \hat{\sigma}_z+ \hbar g \hat{a} \hat{\sigma}_+ + \hbar g^* 
\hat{a}^{\dagger} \hat{\sigma}_- \nonumber \\
&+&\hbar \gamma \hat{\sigma}_+ \hat{a} \big(\hat{b}+\hat{b}^\dagger\big)+ \hbar \gamma^* \hat{\sigma}_- \hat{a}^\dagger 
\big(\hat{b}+\hat{b}^\dagger\big), 
\end{eqnarray}
where we introduced the detuning $\Delta=\omega_{21}-\omega_c$.

\subsection{Dressing the model}
\label{model}

For a pair of bare states with $n$ excitations in the radiation field mode, 
there are two dressed states $\ket{+,n}$ and $\ket{-,n}$. We express these as superpositions of 
the bare states $\ket{1}\ket{n}$ and $\ket{2}\ket{n-1}$ so that
\begin{eqnarray}
  \ket{+,n}&=&\alpha_+(n)\ket{1}\ket{n}+\beta_+(n)\ket{2}\ket{n-1}, \\
 \ket{-,n}&=&\alpha_-(n)\ket{1}\ket{n}+\beta_-(n)\ket{2}\ket{n-1}.
\end{eqnarray}
The eigenvalue equation reads
\begin{eqnarray}
 \Big(\frac{1}{2} \Delta \hat{\sigma}_z&+&  g \hat{a} \hat{\sigma}_+ +  g^* 
\hat{a}^{\dagger} \hat{\sigma}_-\Big) \ket{\pm,n}= \pm \Omega_R(n) \ket{\pm,n}, \nonumber \\
 \Omega_R(n)&=&\sqrt{\Delta^2/4 + |g|^2 n},
\end{eqnarray}
where $\Omega_R(n)$ is the Rabi frequency for $n$ photons, and the coefficients $\alpha_{\pm}(n)$ and $\beta_{\pm}(n)$ are given by
\begin{eqnarray}
 \alpha_-(n)&=&\beta_+(n)=\Big( \frac{\Omega_R(n)+\Delta/2}{2 \Omega_R(n)}\Big)^{1/2}, \\
\alpha_+(n)&=&\Big( \frac{\Omega_R(n)-\Delta/2}{2 \Omega_R(n)}\Big)^{1/2}\ee^{-i \phi}, \\
\beta_-(n)&=&-\Big( \frac{\Omega_R(n)-\Delta/2}{2 \Omega_R(n)}\Big)^{1/2}\ee^{i \phi}.
\end{eqnarray}
We used the ortogonality condition $\bra{+,n} -,n\rangle=0$ and the notation $g=|g|\ee^{i \phi}$.

Now, motivated by the results of the resonant interaction \cite{Bernad}, where a maximally entangled state can be postselected
by a von Neumann measurement, we simplify our model to $\Delta=0$. 
This leads to the following identities
\begin{eqnarray}
  \Big(  g \hat{a} \hat{\sigma}_+ &+&  g^* 
\hat{a}^{\dagger} \hat{\sigma}_-\Big) \ket{\pm,n}= \pm |g|\sqrt{n} \ket{\pm,n}, \nonumber \\
  \Big(  \gamma \hat{a} \hat{\sigma}_+ &+&  \gamma^* 
\hat{a}^{\dagger} \hat{\sigma}_-\Big) \ket{\pm,n}= \pm |\gamma|\sqrt{n} \ket{\pm,n},
\end{eqnarray}
where we used the relation $\gamma/g=|\gamma|/|g|$ supported by the definiton $\gamma=\eta g'(0)$.

We recall the Hamiltonian in \eref{modelI1}, which is block-diagonal in regards to the sectors of 
$\ket{+,n}$ and $\ket{-,n}$
\begin{eqnarray}
 \bra{+,n} \hat{H}_I \ket{+,n}&=&\hbar \omega_t \hat{b}^\dagger \hat{b}+ \hbar |g|\sqrt{n} + \hbar |\gamma| \sqrt{n} 
\big(\hat{b}+\hat{b}^\dagger\big), \nonumber \\
 \bra{-,n} \hat{H}_I \ket{-,n}&=&\hbar \omega_t \hat{b}^\dagger \hat{b}- \hbar |g|\sqrt{n} - \hbar |\gamma| \sqrt{n} 
\big(\hat{b}+\hat{b}^\dagger\big), \nonumber \\
 \bra{+,n} \hat{H}_I \ket{-,n}&=&\bra{-,n} \hat{H}_I \ket{+,n}=0. 
\end{eqnarray}
Let us consider that the centre-of-mass motion state is initially in the ground state. 
We get the following equations by using a general Baker-Hausdorff identity, derived in \ref{BH}:
\begin{eqnarray}
 &&\ee^{-\frac{i}{\hbar} \hat{H}_I t} \ket{+,n,0}=\ee^{-i \omega_t \hat{b}^\dagger \hat{b} t-i |\gamma| \sqrt{n} 
\big(\hat{b}+\hat{b}^\dagger\big)t -i|g|\sqrt{n} t} \ket{+,n,0} \nonumber \\
&&=\ee^{-i \omega_t \hat{b}^\dagger \hat{b} t-i |\gamma| \sqrt{n} 
\big(\hat{b}+\hat{b}^\dagger\big)t -i |g|\sqrt{n} t}\ee^{i \omega_t \hat{b}^\dagger \hat{b} t}\ee^{-i \omega_t \hat{b}^\dagger \hat{b} t}
\ket{+,n,0} \nonumber \\
&&=\ee^{i\Phi_n(t)-i|g|\sqrt{n} t} \ee^{\alpha_n(t)\hat{b}^\dagger-\alpha^*_n(t)\hat{b}} \ee^{-i \omega_t \hat{b}^\dagger \hat{b} t}  
\ket{+,n,0} \nonumber \\
&&=\ee^{i\Phi_n(t)-i|g|\sqrt{n} t} 
\ket{+,n,-\alpha_n(t)}, \label{evolution}
\end{eqnarray}
and for the state $\ket{-,n,0}$
\begin{equation}
 \ee^{-\frac{i}{\hbar} \hat{H}_I t} \ket{-,n,0}=\ee^{i\Phi_n(t)+i|g|\sqrt{n} t} 
\ket{-,n,\alpha_n(t)},
\end{equation}
where we introduced
\begin{eqnarray}
 \alpha_n(t)&=&\frac{|\gamma|\sqrt{n}}{\omega_t}\left(1-\ee^{-i\omega_t t}\right), \\
 \Phi_n(t)&=&\frac{|\gamma|^2n}{\omega^2_t}\left(\omega_t t-\sin(\omega_t t)\right).
\label{displacement}
\end{eqnarray}

\subsection{Hybrid quantum repeater setup}
We consider two spatially separated three-level systems $A$ and $B$,
with internal energy eigenstates  $\ket{0}_i, \ket{1}_i$, and $\ket{2}_i$ ($i\in \{A,B\}$).
The states $\ket{0}_i, \ket{1}_i$ serve as the qubits which are going to be entangled. The interaction between the states
$|1\rangle_i$ and $|2\rangle_i$ is given by the model of a trapped material system, discussed in chapter \ref{modelH}.

Our main purpose is to investigate the influence of the centre-of-mass motion in the entanglement creation by a 
minimum-error POVM (positive operator-valued measure) measurement. Furthermore, we consider a Ramsey-type interaction scenario 
as illustrated in figure \ref{fig-1} in order to entangle the two material quantum systems $A$ and $B$.
In a first step the single-mode radiation field of cavity $A$ interacts with the three-level system $A$ during a time
interval of duration $\tau$ by a Stark-switching procedure. This is followed by a perfect state transfer
between cavities $A$ and $B$ by an optical fibre \cite{Bernad}. The whole system evolves freely for a time $T$ during the 
propagation of the optical radiation field from system $A$ to system $B$. In the last and final step the single-mode 
radiation field in cavity $B$
interacts with system $B$ for a time $\tau$, again by employing a Stark-switching procedure. The whole process takes a time $t=2\tau+T$ 
where the time for the transfers and for the propagation in the optical fibre is given jointly by $T$. The complete procedure could 
take a long duration, hence to avoid the effects of spontaneous decay from the states $\ket{2}_i$, one may transfer the information stored
on these levels to radiatively stable levels immediately after each interaction ocurred between the material qubit system and 
local single-mode field. 

Initially, the system is assumed to be prepared in a product state
\begin{eqnarray}
\ket{\Psi(t=0)}&&=\frac{\ket{0}_A+\ket{1}_A}{\sqrt{2}}\otimes\frac{\ket{0}_B+\ket{1}_B}{\sqrt{2}}\otimes \\
&&\otimes \ket{\alpha}_{c,A} \otimes 
\ket{0}_{t,A}\otimes \ket{0}_{t,B}
 \otimes \ket{0}_f \otimes \ket{0}_{c,B}. \nonumber
\end{eqnarray}
This state can be prepared by local operations and by a laser cooling procedure. The cooling procedure allows the preparation
of the centre-of-mass motion in the ground state $\ket{0}_{t,i}$ ($i\in \{A,B\}$). The single-mode
radiation field in cavity $A$ is assumed to be prepared in the coherent state $\ket{\alpha}_{c,A}$ with the mean photon number
$\overline{n} = |\alpha|^2$. The states of the optical fibre and of the cavity $B$ are considered to be prepared
in vacuum, i.e., $\ket{0}_f$ and $\ket{0}_{c,B}$.
 
The dynamics in each cavity is described by the Hamiltonian
\begin{eqnarray}
\hat{H}_i=\hat{H}_0+\hbar g \hat{a}_i \hat{\sigma}^i_+  +  \hbar g^* \hat{a}^{\dagger}_i \hat{\sigma}^i_- 
+\hbar \big(\gamma \hat{\sigma}^i_+ \hat{a}_i+\gamma^* \hat{\sigma}^i_- \hat{a}^\dagger_i\big) \big(\hat{b}_i+\hat{b}^\dagger_i\big),
\end{eqnarray}
with $i\in \{A,B\}$ and
\begin{eqnarray}
\hat{H}_0&=&\sum_{i=A,B} \left(\hbar \omega_t \hat{b}^\dagger_i \hat{b}_i+\hbar \omega_0 \ket{0}_i \bra{0}_i+ 
\frac{1}{2}\hbar \omega_{21} \hat{\sigma}^i_z 
+\hbar \omega_c \hat{a}^\dagger_i \hat{a}_i\right)  \nonumber \\
&+& \sum_j \omega_j \hat{a}^\dagger_{f,j} \hat{a}_{f,j},
\end{eqnarray}
where we considered completely similar trapped systems, so we have symmetric couplings $g = g_A = g_B $ and $\gamma = \gamma_A =\gamma_B$. 
$\hat{a}_A$, $\hat{a}_B$ and $\hat{a}_{f,j}$ ($\hat{a}^\dagger_A$, $\hat{a}^\dagger_B$ and $\hat{a}^\dagger_{f,j}$) 
are the annihilation (creation) operators of the modes of the cavities $A$, $B$ and of the optical fibre. 
The frequency of the cavity modes is set to be the same $\omega_c$. It is assumed that only a single transverse and many 
longitudinal modes of the optical fibre are relevant in the propagation process. The frequencies of the fibre modes $\omega_j$ 
are defined by the relation $\omega_j=2 \pi c j/L$ with integer values of $j$ and with the length of the optical fibre $L$.   

The modes of the optical fibre which resonantly couple to the mode of cavity $A$ and $B$ are assumed to form a frequency band 
$(\omega_c-\delta \omega,\omega_c+\delta \omega)$. In the rotating wave approximation $\delta \omega \ll \omega_c$ the 
coupling between the single mode cavities and the
optical fibre modes is described by the Hamiltonian
\begin{eqnarray}
\hat{H}_i= \hbar \omega_c \hat{a}^\dagger_i\hat{a}_i +
\sum_{j} \hbar \omega_j \hat{a}^\dagger_{f,j}\hat{a}_{f,j} 
+ \sum_{j} \left( \kappa_{i,j} 
\hat{a}^\dagger_{f,j} \hat{a}_i+ \kappa^*_{i,j} 
\hat{a}^\dagger_i \hat{a}_i\right), 
\end{eqnarray}
where $\kappa_{i,j}$($i \in\{A,B\}$) describes the coupling between the cavity modes and the $j$th mode of the optical fibre. 
We consider the following conditions (see Ref. \cite{Bernad})
\begin{eqnarray}
\kappa_{B,j} &=& \kappa_{A,j}^* = \mid \kappa_{A,j}\mid e^{-i\varphi_j},\nonumber\\
e^{2i\varphi_j} &=& \frac{\hbar \omega_j - \hbar \omega_c + i\hbar \Gamma_A/2}
{\hbar \omega_j - \hbar \omega - i\hbar \Gamma_A/2}
\end{eqnarray}
with $\Gamma_A$ being the decay rate of cavity $A$ and assume a long fibre such that $L \gg c/\Gamma_A$. These conditions ensure that
the leakage out of cavity $A$ into the optical fibre and also out of the optical fibre into cavity $B$ is much shorter than the propagation
of the radiation field in the fibre and a perfect photonic state transfer between the two cavities is realized. 

Now, considering the resonant matter-field interaction $\Delta=\omega_{21}-\omega_c=0$, the quantum state $\ket{\Psi(t)}$ 
in this Ramsey-type interaction sequence results:
\begin{eqnarray}
\label{Psi}
&&\ket{\Psi(t)} =\Big(
\frac{1}{2} \ket{0}_A \ket{0}_B \ket{\alpha\ee^{-i\omega t}}_{c,B} \ket{0}_{t,A} \ket{0}_{t,B} \ee^{-i\Phi_{00}}\nonumber \\
&&+\ket{g_{10}(t)} \ket{1}_A \ket{0}_B \ee^{-i\Phi_{10}} + 
\ket{g_{01}(t)} \ket{0}_A \ket{1}_B \ee^{-i\Phi_{01}} + 
\ket{g_{20}(t)} \ket{2}_A \ket{0}_B \ee^{-i\Phi_{20}} \nonumber \\
&&+\ket{g_{02}(t)} \ket{0}_A \ket{2}_B \ee^{-i\Phi_{02}} +
\ket{g_{11}(t)} \ket{1}_A \ket{1}_B \ee^{-i\Phi_{11}} +
\ket{g_{12}(t)} \ket{1}_A \ket{2}_B \ee^{-i\Phi_{12}}   \nonumber \\
&&+\ket{g_{21}(t)} \ket{2}_A \ket{1}_B \ee^{-i\Phi_{21}}+ \ket{g_{22}(t)} \ket{2}_A \ket{2}_B \ee^{-i\Phi_{22}}\Big)\otimes \ket{0}_{c,A} \otimes \ket{0}_f,
\end{eqnarray}
with the phase factors
\begin{eqnarray}
 \Phi_{00}=2 \omega_0 t, \,\,\,\, \Phi_{10}=\Phi_{01}=\omega_0 t  - \frac{\omega_c}{2} (T + 2\tau),
\end{eqnarray}
\begin{eqnarray}
 \Phi_{20}=\omega_0 t + \frac{\omega_c}{2} (T+2\tau), \,\,\,\, \Phi_{02}=\omega_0 t  + \frac{\omega_c}{2} (T+2\tau), 
\end{eqnarray}
\begin{eqnarray}
\Phi_{11}=-\omega_c (2\tau+T), \,\,\,\, \Phi_{12}=\Phi_{21}=0, \,\,\,\, \Phi_{22}=\omega_c t.
\end{eqnarray}

The unnormalized states $\ket{g_{ij}(t)}$ ($i,j=0,1,2$) entering \eref{Psi} describe the state of the radiation field 
in cavity $B$ and the states of the centre-of-mass motions for both trapped systems. We show the detailed structure of these states in 
\ref{Calc}.

The quantum state of \eref{Psi} yields a complete description of the interaction between the trapped systems $A$ and $B$ 
and the optical radiation fields in the  case of resonant interaction $\omega_{21}=\omega_c$, 
i.e. neglecting all other decoherence sources except for the centre-of-mass motion.
It can easily be shown that the overlap between $\ket{g_{01}(t)}$ and $\ket{g_{10}(t)}$ is the highest compared to all other overlap
combinations, and the probability of projecting onto $\ket{0}_A \ket{1}_B$ or $\ket{1}_A \ket{0}_B$ is the same. This means that 
the most promising scenario is to project onto the qubit subspace spanned by $\ket{0}_A\ket{1}_B$ and $\ket{1}_A\ket{0}_B$, 
and a POVM measurement on the single-mode field of cavity $B$ 
could prepare a material Bell state $\ket{\Psi^+} = (\ket{0}_A \ket{1}_B + \ket{1}_A \ket{0}_B)/\sqrt{2}$ by photonic postselection.  

Let us start from the pure quantum state $|\Psi(t)\rangle$ of \eref{Psi} and the field state $\hat{\rho}_F(t)$ appearing in the photon
detector that is obtained by tracing out the material degrees of freedom, the radiation field state in cavity $A$ and 
the radiation field states of the optical fibre (both of these radiation field states being in the vacuum state),
\begin{eqnarray}
\hat{\rho}_F(t) &=& \bra{0}_{c,A}\bra{0}_f{\rm Tr}_{A,B,traps}\{ |\Psi(t)\rangle \langle \Psi(t)| \}\ket{0}_f \ket{0}_{c,A} \nonumber \\
   &=&   p \hat{\rho}_1 + (1-p) \hat{\rho}_2,  
\label{fieldstate}
\end{eqnarray}
with the unnormalized field states
\begin{eqnarray}
 &&p\hat{\rho}_1=\sum^{\infty}_{n,m=0} \Big(a_{10}(n,m)+a_{10}(n,m)\Big) \ket{n}_{c,B}\bra{m}_{c,B}, \\
&&(1-p)\hat{\rho}_2=\frac{1}{4} \ket{\alpha \ee^{-i \omega_c t}}_{c,B}\bra{\alpha \ee^{-i \omega_c t}}_{c,B}+  \\
&&+\sum^2_{i,j=0}\sum^{\infty}_{n,m=0} a_{ij}(n,m)\ket{n}_{c,B}\bra{m}_{c,B},\,\,(i,j)\neq (1,0) , (0,1) \nonumber.
\end{eqnarray}
The coefficients $a_{ij}(n,m)$ are given in \ref{Calc} and the normalization factor is
\begin{eqnarray} \label{eq:probability}
 p=\frac{1}{4} \ee^{-|\alpha|^2}\sum^{\infty}_{n=0} \frac{|\alpha|^{2n}}{n!}\Big[1+\cos\big(2|g|\sqrt{n}\tau\big)
 \ee^{-\frac{4n|\gamma|^2}{\omega^2_t}
\Big(1-\cos(\omega_t \tau)\Big)} \Big].
\end{eqnarray}
The quantum state $\hat{\rho}_2$ is a mixed state, furthermore $\hat{\rho}_1$ and $\hat{\rho}_2$ are not orthogonal,
therefore we discard the strategy of unambiguous discrimination, which has difficulties treating mixed states \cite{Chefles,Chefles2,Bruss}.

In order to optimize the fidelity and success probability for a postselected 
entangled Bell state $|\Psi^+\rangle$ it is necessary to perform a minimum-error POVM measurement on the 
optical radiation field, since the smallest possible failure probability in unambiguous discrimination is at least twice as large
as the smallest error probability in minimum-error discrimination for an arbitrary mixed state \cite{Bergou}.

\begin{figure}[t]
\begin{center}
\includegraphics[width=7cm]{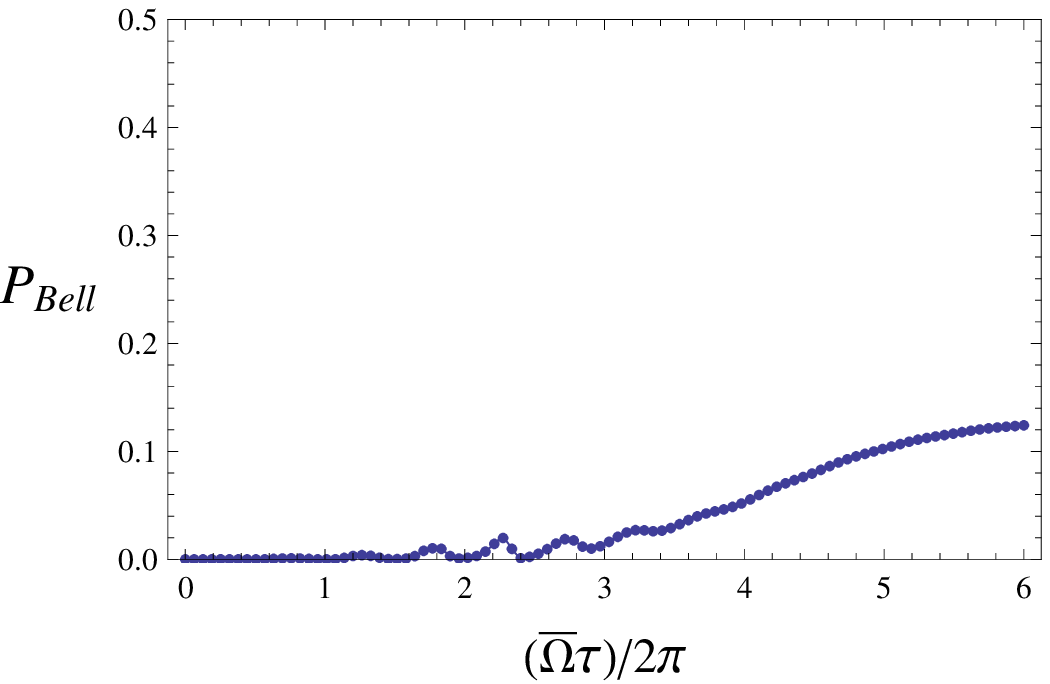}
\includegraphics[width=7cm]{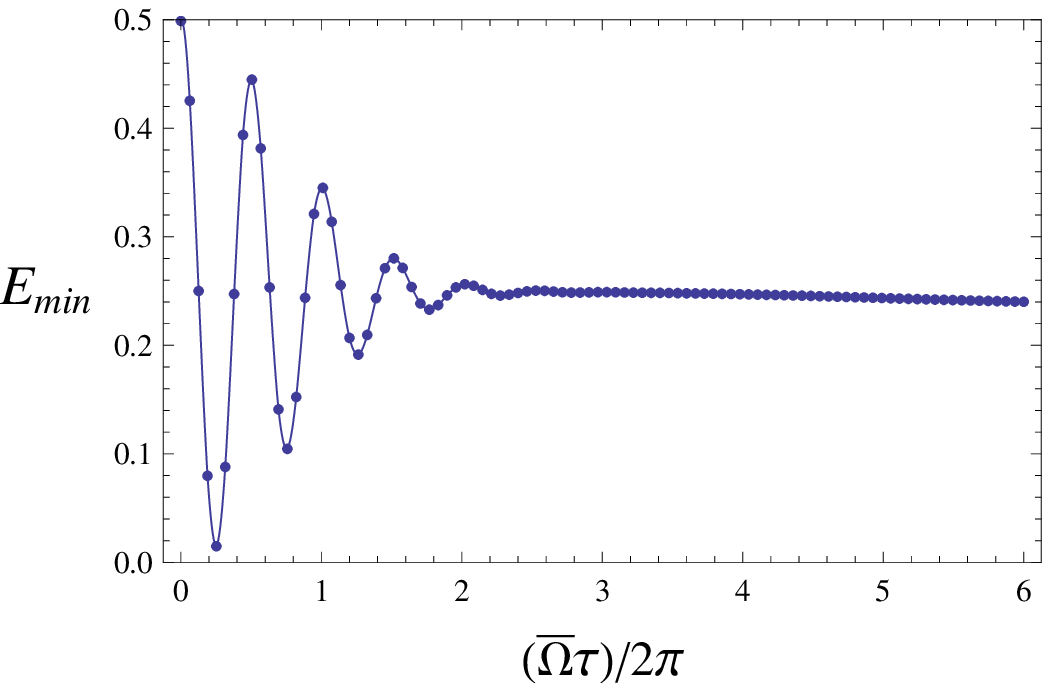}
\includegraphics[width=7cm]{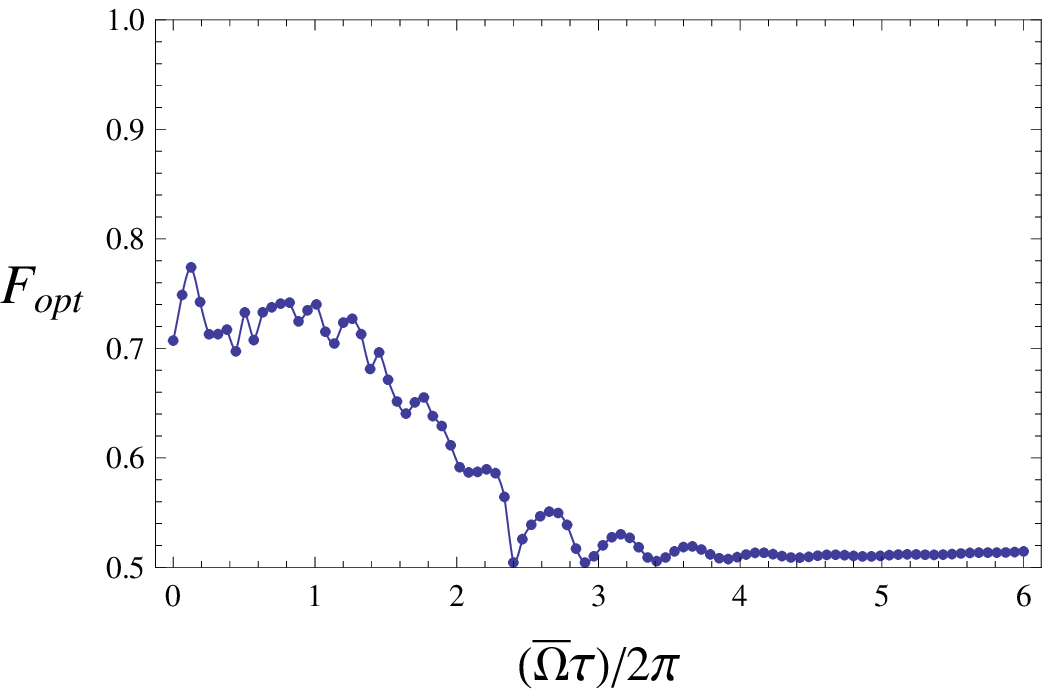}
\caption{\label{fig-2} Success probability, minimum-error probability and the optimal fidelity for a small trap frequency, 
i.e. $\frac{|\gamma|}{|g|}=0.1$ and $\frac{\omega_t}{|g|\sqrt{\bar{n}}}=0.1$. The figures show a low fidelity postselected pair
with a $25\%$ percent success probability and a $50\%$ error probability in the time interval of the collapse phenomenon. 
The average number of photons is $\bar{n}=10^2$ and 
$\bar{\Omega}=|g|\sqrt{\bar{n}}$.} 
\end{center}
\end{figure}

\begin{figure}[t]
\begin{center}
\includegraphics[width=7cm]{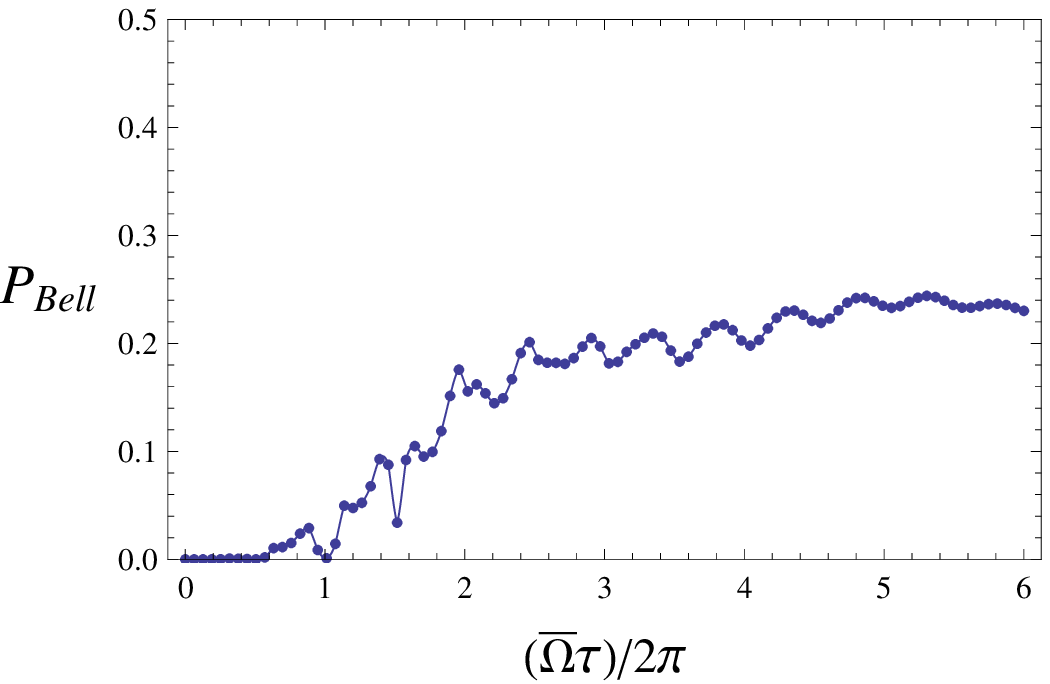}
\includegraphics[width=7cm]{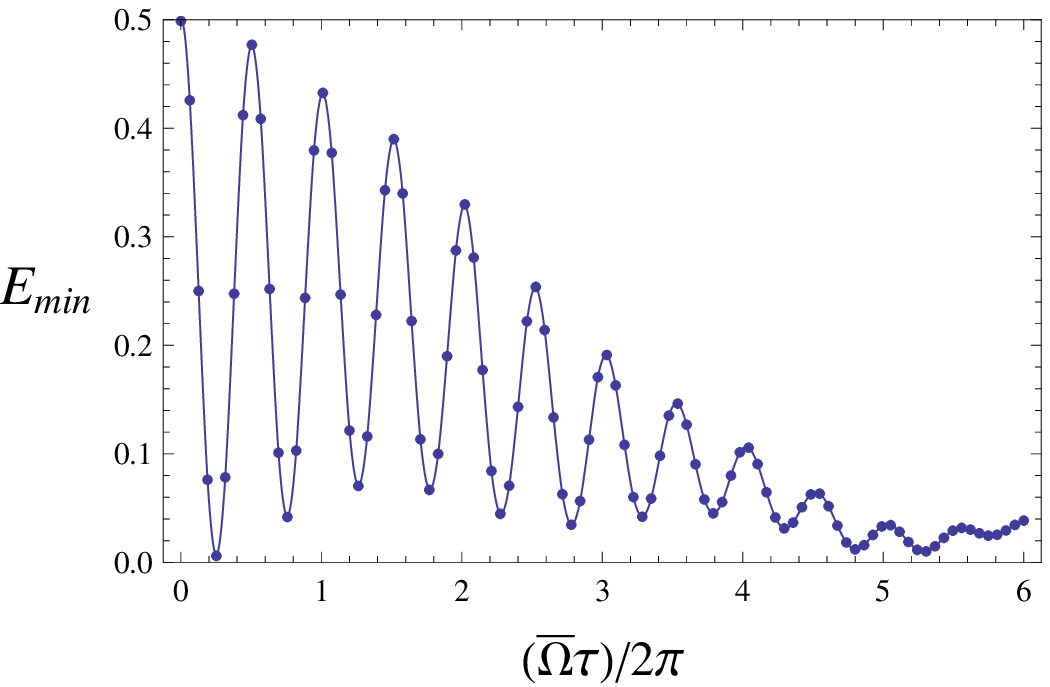}
\includegraphics[width=7cm]{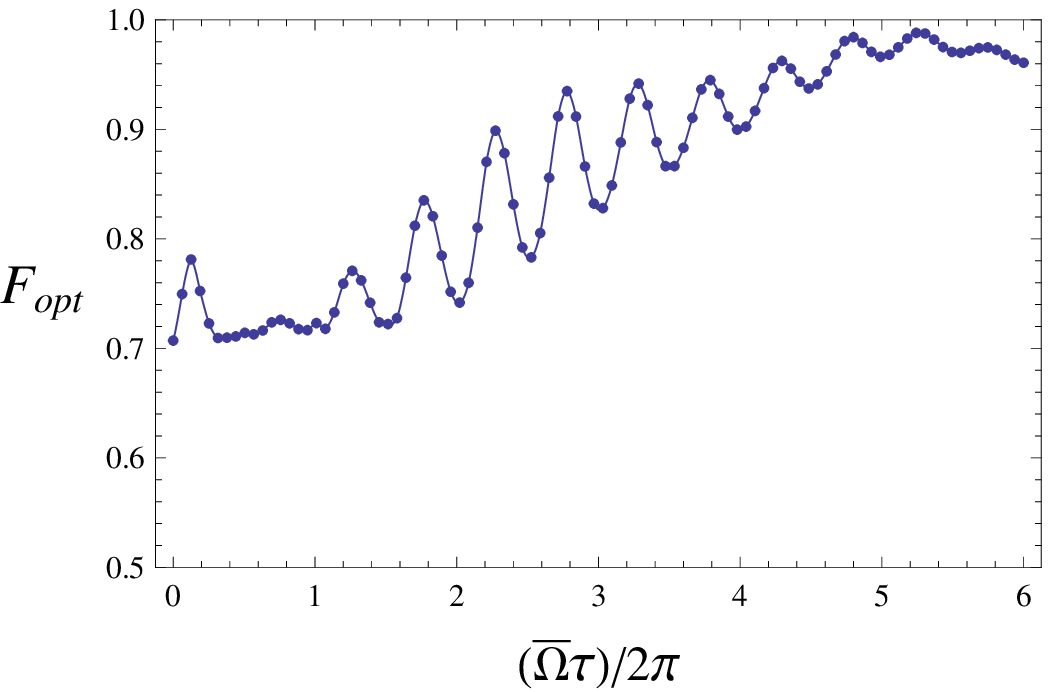}
\caption{\label{fig-3} Success probability, minimum-error probability and the optimal fidelity for a large trap frequency, 
i.e. $\frac{|\gamma|}{|g|}=0.05$ and $\frac{\omega_t}{|g|\sqrt{\bar{n}}}=0.4$. The figures more closely resemble the result found for 
the motionless model. The average number of photons is $\bar{n}=10^2$ and 
$\bar{\Omega}=|g|\sqrt{\bar{n}}$.} 
\end{center}
\end{figure}

The measurement is performed on the field state $\hat{\rho}_F(t)$ and has two possible outcomes $\lambda=0,1$. 
The measurement outcome $\lambda=1$ corresponds to a projection onto the field state $\hat{\rho}_1$
and the measurement outcome $\lambda=0$
corresponds to a projection onto the field state  $\hat{\rho}_2$.
We denote the positive operators of these two measurements by
$\hat{T}\geq 0 $ and $I - \hat{T}$, with $I$ being the unit operator on the Hilbert space of the single-mode
radiation field. The problem in minimum-error state discrimination is to examine the tradeoff between the
two error probabilities ${\rm Tr}\{\hat{T} \hat{\rho}_2\}$ and  ${\rm Tr}\{(I - \hat{T}) \hat{\rho}_1\}$, and
the positive operator $\hat{T}$ has to be determined in such
a way that for a given {\it a priori} probability $p$ from \eref{eq:probability} the error probability
\begin{eqnarray}
E &=& p {\rm Tr}\{(I - \hat{T}) \hat{\rho}_1\} + (1-p){\rm Tr}\{\hat{T} \hat{\rho}_2\}
\label{E} 
\end{eqnarray}
is minimal. Diagonalizing the Hermitian operator $\hat{X}:= p\hat{\rho}_1 - (1-p)\hat{\rho}_2 $, which results in
$\hat{X} = \sum_{x} x \ket{x}_{c,B}\bra{x}_{c,B} $, 
the solution of this optimization problem is given by the projection operator \cite{Helstrom,Holevo,Hayashi}
\begin{eqnarray}
\hat{T} &=& 
\sum_{x\geq 0} \ket{x}_{c,B}\bra{x}_{c,B}
\label{optimalPOVM}
\end{eqnarray}
which projects onto eigenstates of the operator $\hat{X}$ belonging to non-negative eigenvalues.
By inserting the optimum detection operator $\hat{T}$ into \eref{E} the minimum error probability $E_{min}$ 
is found to be (see Ref. \cite{Hayashi})
\begin{eqnarray}
E_{min} = \frac{1}{2}\left( 1 - ||p \hat{\rho}_1 - (1-p) \hat{\rho}_2||_1 \right),
\label{EMin}
\end{eqnarray}
with $||.||_1$ being the trace norm. The probability $P_{Bell}$ that the minimum-error POVM measurement
prepares the spatially separated quantum systems $A$ and $B$ in the Bell state $|\Psi^+\rangle$ of the internal states is given by
\begin{eqnarray}
P_{Bell} = p{\rm Tr}_{field}\{\hat{\rho}_1 \hat{T}\}.
\label{PBell}
\end{eqnarray} 
After a successful minimum-error POVM measurement, the joint internal state of both quantum systems $A$ and $B$ 
is given by
\begin{eqnarray}
\hat{\rho}_{AB}(t) &=&
\frac{
{\rm Tr}_{fields,traps}
\{
|\Psi(t) \rangle \langle \Psi(t)|
\hat{T}
\}
}{
{\rm Tr}_{A,B,fields,traps}
\{
|\Psi(t) \rangle \langle \Psi(t)|
\hat{T}
\}
}. 
\end{eqnarray}
Thereby, the fidelity $F_{opt}$ of an optimally prepared Bell pair which is postselected by a measurement result with value $\lambda=1$ 
is given by
\begin{eqnarray}
F_{opt} &=&
\sqrt{\langle \Psi^+| \hat{\rho}_{AB}(t) |\Psi^+\rangle}.
\label{OptFid}
\end{eqnarray}
In the following these quantities are calculated numerically. We concentrate on the case of large numbers of photons, 
i.e. $\overline{n} = 10^2$ and on values of the interaction times $\tau$ where the collapse phenomenon occurs. In figures \ref{fig-2} and
\ref{fig-3} numerical results are presented to reveal the postselection by a minimum-error POVM measurement on the optical radiation 
field in order to prepare a Bell state $|\Psi^+\rangle$. These numerical results are based on the quantum state of
\eref{Psi}. The minimum-error POVM measurement is determined according to \eref{optimalPOVM}. This optimal POVM measurement 
depends on the  following electrodynamical interaction parameters: the
interaction time $\tau$, the mean photon number $\overline{n}$, the trap frequency $\omega_t$, 
the resonant  Rabi frequency $\overline{\Omega}=|g|\sqrt{\bar{n}}$ and the 
strength of the coupling to the centre-of-mass motion $\gamma$, which depends on the
trap frequency $\omega_t$ in the Lamb-Dicke regime like $\gamma \sim \omega_t^{-1/2}$ (see \eref{eta}).

In the ideal case \cite{Bernad} it was found that a perfectly entangled state can be prepared with probability $25\%$. This situation
occurs during the collapse phenomenon of the Jaynes-Cummings-Paul model. In our model we found that these results 
are very sensitive to the trap frequency, see figures \ref{fig-2} and \ref{fig-3}. If we consider a trap frequency for which the 
results resemble the ideal case, then for a four times smaller trap frequency the best fidelity achieved is $F_{opt}=0.5$ with a 
probability of $11\%$. These results are consistent with the expectation that the centre-of-mass 
motion introduces a significant amount of decoherence in the system. 
This decoherence prohibits the creation of high-fidelity pairs. In order to increase the characteristic quantities we must increase 
the frequency of the trap, see figure \ref{fig-3}. The increase of the trap frequency corresponds to a steeper harmonic potential, which is 
reducing the centre-of-mass motion. However, in the case of already built experimental apparatus the eigenfrequency of the trap 
can not be manipulated at will. Therefore, the preparation of high-fidelity Bell states is limited by the 
centre-of-mass motion even if the postselection is performed by minimum-error POVM measurements.


\section{Decoupling the centre-of-mass motion}
\label{decoupling}

In this section we look at dynamical decoupling techniques to suppress the unwanted interaction between the centre-of-mass 
motion on the one hand and the the radiation field and the internal states on the other hand. Dynamical decoupling in finite dimensional systems was introduced by Viola et al. \cite{Viola1} and 
subsequently developed by different authors \cite{Viola2,Zanardi}  as a general approach to fight decoherence in open quantum systems 
by repeatedly 
acting on the system in a controlled way such that the influence of unwanted environmental couplings cancel out. 
As an implementation of the original idea a dynamical decoupling approach was suggested by Vitali and Tombesi \cite{Vitali} for 
two coupled harmonic oscillators within the rotating wave approximation. 

While these papers formalized 
and generalized the idea, the principle of such control actions has been known and used even earlier in the NMR community \cite{Haeberlen}, 
and the spin-echo effect \cite{hahn} is the first and probably most well-known application of the concept. The aforementioned papers focus on finite or countably infinite 
dimensional systems interacting with an arbitrary environment, however our requirements are slightly different. Our goal is to protect the 
subspace consisting of the finite internal three-level state and the radiation field, but in contrast to usual applications of 
dynamical decoupling we cannot act on the internal states of the qubits because that would decouple not only the unwanted interaction with the 
centre-of-mass motion, 
but also the required interaction with the radiation field. Instead, we have to act on our environment, which is the harmonic oscillator 
space of the centre-of-mass motion. Similar to the work of Vitali and Tombesi \cite{Vitali} we have to apply a sequence of operations on the motion's harmonic oscillator, but our 
situation is more complicated since we have to ensure that we do not interfere with the interaction between 
the internal states and the radiation field. 

The basic idea is to use a decoupling scheme locally in each cavity during the characteristic time $\tau$ of the collapse phenomena. For this purpose we need to
use the dynamical result of section \ref{The model}. The time evolution derived in \eref{evolution} shows that 
the oscillator states of the centre-of-mass motion and the 
joint states of the radiation field and of the three level system are entangled. This entanglement is detrimental to the quantum repeater 
and needs to be eliminated, if possible. A first step is to observe that the coherent state displacement $\alpha_n(t)$ in the oscillator 
space oscillates with the 
trap frequency $\omega_t$ and vanishes for times $t = k \frac{2\pi}{\omega_t}, k \in [0,1,2,\dots]$ for all $n$. Since this 
oscillation is faster than the interaction time $\tau$, one should try and choose $\tau = k \frac{2\pi}{\omega_t}$ while ensuring 
that $\tau$ remains in the immediate vicinity of the occurence of the collapse phenomenon. Doing so 
ensures that the oscillator state is separable at the end of the interaction.
However, the motion-field interaction still introduces relative phase shifts $e^{i\Phi_n(t)}$ which do not cancel so easily. 
To have all the phases vanish simultaneously, one would require that $t = m \frac{2\pi \omega_t}{|\gamma|^2}, m \in [0,1,2,\dots]$, in 
addition to $t = k \frac{2\pi}{\omega_t}$. This is fulfilled if $\frac{k}{m} = \frac{\omega_t^2}{|\gamma|^2}$ which will generally 
lead to very large $t \gg \tau$ and is hardly achievable in the constraints of this setup. 

\subsection{Finding a decoupling scheme}

We now focus on the suppression of the coupling between the centre-of-mass motion and the rest of the system in the Hamiltonian 
\begin{eqnarray}
\label{modelI}
\hat{H}_I&=& \hbar \omega_t \hat{b}^\dagger \hat{b}+ \hbar g \hat{a} \hat{\sigma}_+ + \hbar g^* 
\hat{a}^{\dagger} \hat{\sigma}_- \nonumber \\
&+&\hbar \gamma \hat{\sigma}_+ \hat{a} \big(\hat{b}+\hat{b}^\dagger\big)+ \hbar \gamma^* \hat{\sigma}_- \hat{a}^\dagger 
\big(\hat{b}+\hat{b}^\dagger\big) 
\end{eqnarray}
by a dynamical decoupling scheme. To that end we assume that we are able to apply instantaneously (\emph{bang-bang} control \cite{Viola1}) 
a single unitary operator $\hat{p}$ 
to the motion subspace repeatedly with a frequency of $\frac{1}{\Delta t}$. In general dynamical decoupling methods allow for the application
 of several different unitary operators $\hat{p}_i$, but we will see shortly that a single operator is sufficient in our case.
The resulting time evolution after application of $N$ pulses at time $t=N \Delta t$ is
\begin{equation}
\hat{U}_N(t) = \left( \hat{p} e^{-\frac{i}{\hbar} \hat{H}_I \frac{t}{N}} \right)^N .
\end{equation}
By calculating the time derivative of $\hat{U}_N(t)$ we can define an average Hamiltonian $\hat{H}_N(t)$ which generates the same time
 evolution:
\begin{eqnarray}
 \frac{d}{dt}\hat{U}_N(t)&=&-\frac{i}{N} \sum^{N-1}_{k=0} \left( \hat{p} e^{-\frac{i}{\hbar} \hat{H}_I \frac{t}{N}} \right)^k \hat{p} 
\hat{H}_I \hat{p}^\dagger \left[ \left( \hat{p} e^{-\frac{i}{\hbar} \hat{H}_I \frac{t}{N}} \right)^\dagger \right]^k \hat{U}_N(t) \nonumber \\
&\equiv &-i\hat{H}_N(t) \hat{U}_N(t). 
\end{eqnarray}
In order for our decoupling scheme to have the desired effect, the average Hamiltonian $\hat{H}_N$ should be equal to $\hat{H}_I$
 minus the interaction with the centre-of-mass motion. Although perfect equality is generally not possible, we will try to get as close as
 we can.

To find suitable candidates for the operator $\hat{p}$ we regard the limit of continuous control, i.e. $N \to \infty$. We are going to 
derive the generator of the time evolution in this limit by following the method given in the work of Facchi et. al. \cite{facchi}. The 
limiting unitary evolution 
\begin{equation}
\hat{\mathcal{U}}(t)  = \lim_{N \to \infty} \hat{U}_N(t)
\end{equation}
satisfies the equation
\begin{eqnarray} 
\frac{d}{dt} \hat{\mathcal{U}}(t) &=& -i \hat{\mathcal H} \hat{\mathcal{U}}(t) , \nonumber \\
\hat{\mathcal H} &=& \lim_{N \to \infty} \frac{1}{N} \sum_{k=0}^{N-1} \hat{p}^{k+1} \hat{H}_I \left(\hat{p}^\dagger \right)^{k+1}. \label{eq:H_limit}
\end{eqnarray}
Let us begin by outlining our goal. We are looking for those $\hat{p}$ which satisfy the following equation
\begin{equation}
 \hat{\mathcal H} = \hat{H}_\text{id} \equiv \hbar \omega_t \hat{b}^\dagger \hat{b}+ \hbar g \hat{a} \hat{\sigma}_+ + \hbar g^* 
\hat{a}^{\dagger} \hat{\sigma}_- ,
\end{equation}
such that in the limit $N \to \infty$ $\hat{H}_N$ approaches the ideal Hamiltonian.
Since $\hat{p}$ acts on the subspace of the centre-of-mass motion and $\hat{H}_I-\hat{H}_\text{id}\sim \hat{b}+\hat{b}^\dagger$, 
it turns out that
the solution  is to choose $\hat{p}$ as a diagonal operator in the oscillator eigenstates,
\begin{equation}
\label{pform}
 \hat{p}=\sum_{n=0}^\infty e^{-i\lambda_n} \ket{n}_t\bra{n}_t\otimes \hat{I}_c
\otimes \hat{I}_3,
\end{equation}
where $\ket{n}_t$ ($n\in {\mathbb N}$) is the number state representation of the centre-of-mass motion, $\hat{I}_c$ is the identity 
operator on the 
Fock space of the radiation field and $\hat{I}_3$ stands for the three dimensional identity matrix. $\hat{b}^\dagger \hat{b}$ is a diagonal 
operator, which means that it commutes with $\hat{p}$.
Facchi et. al. \cite{facchi} studied the effects of decoupling operators in the form of  \eref{pform} with non-degenerate spectra,
 i.e. $\lambda_n \neq \lambda_m \pmod{2\pi} \text{ for } n \neq m$, but we choose not to make this restriction.
Inserting \eref{pform} into \eref{eq:H_limit} we find:
\begin{eqnarray}
\hat{\mathcal H} &=& \hat{H}_\text{id} + \lim_{N\to \infty} \frac{1}{N} \sum_{k=1}^{N} \hat{p}^{k} (\hat{H}_I - \hat{H}_\text{id}) (\hat{p}^\dag)^{k} \nonumber \\
&=& \hat{H}_\text{id} + \hbar \left(\gamma \hat{\sigma}_+ \hat{a}+ \gamma^* \hat{\sigma}_- \hat{a}^\dagger \right) \lim_{N\to \infty} \sum_{n,m=0}^\infty \nonumber \\
&& \left( \frac{1}{N} \sum_{k=1}^N e^{-i(\lambda_n - \lambda_m) k} \right) \ket{n}_t\bra{n}_t (\hat{b}+\hat{b}^\dag) \ket{m}_t\bra{m}_t \label{eq:intermediary} \\
&=& \hat{H}_\text{id} + \hbar \left(\gamma \hat{\sigma}_+ \hat{a}+ \gamma^* \hat{\sigma}_- \hat{a}^\dagger \right) \sum_{\substack{\lambda_n=\lambda_{n+1} \\ \pmod{2\pi}}} \sqrt{n+1}   \nonumber \\
&& \times \left( \ket{n}_t \bra{n+1}_t + \ket{n+1}_t \bra{n}_t \right)
\end{eqnarray}
The limit of $N\to \infty$ eliminates all pairs of the sum in \eref{eq:intermediary} where $\lambda_n \neq \lambda_m \pmod{2\pi}$ ($n \neq m$).
 Of the remaining pairs only direct neighbours contribute due to the ladder operators $\hat{b}$ and $\hat{b}^\dag$. Therefore, in order for $\hat{\mathcal H}$ 
to be equal to $\hat{H}_\text{id}$, we require that $\lambda_n \neq \lambda_{n+1} \pmod{2\pi}$ for any $n$. Aside from this restriction, our derivation allows for degenerate $\lambda$ values in contrast to the result of 
Facchi et. al. \cite{facchi}. Similar calculations reveal that interactions of 
odd power $(\hat{b}+\hat{b}^\dagger)^j$ vanish if $\lambda_n \neq \lambda_{n+j} \left(\text{mod} \,\,2\pi \right)$.

In the limit of continuous control $N \to \infty$ we found a class of unitary operations which have the form given in equation 
\eref{pform} with the condition that
any two neighbors $\lambda_n$ and $\lambda_{n+1}$ are not allowed to be in the same $2 \pi$ modulo class. While this concludes the search
from a mathematical viewpoint, in the next section we will look at actual unitary operators that fulfill these conditions and look at how
they might be implemented experimentally.

\subsection{Suitable decoupling operators and physical implementation}
There is one particular choice for the decoupling operator $\hat{p}$ which fulfills the conditions 
$\lambda_n \neq \lambda_{n+j} \left(\text{mod} \,\,2\pi \right)$ for all odd $j$. 
That is the parity operator 
\begin{equation}
\hat{\mathcal{P}} = \sum_{n=0}^\infty (-1)^n \ket{n}_t\bra{n}_t,
\end{equation}
whose $\lambda_n$ are $0, \pi, 2\pi, 3\pi, \dots$
This choice of decoupling operator has already been proposed by Vitali and Tombesi \cite{Vitali} for the case of two 
harmonic oscillators interacting in the rotating wave
approximation. 

The parity operator can be written in terms of the number operator $\hat{b}^\dagger\hat{b}$ as
 $\hat{\mathcal{P}} = \ee^{-i \pi \hat{b}^\dagger\hat{b}}$. If we replace $\pi$ by an arbitrary phase $\varphi \in (0,\pi)$, 
then we get a more general class of decoupling operators
\begin{equation}
\hat{p} = \ee^{-i \varphi \hat{b}^\dagger\hat{b}}
\end{equation}
with $\lambda_n$ values of $0, \varphi, 2\varphi, 3\varphi, \dots$
Therefore they still fulfill the necessary condition $\lambda_n \neq \lambda_{n+1} \left(\text{mod} \,\,2\pi \right)$, 
although they may not fulfill the condition for arbitrary odd $j$ as the parity operator does and thus may not decouple higher orders $(\hat{b}+\hat{b}^\dag)^j$, $j=3,5,\dots$.
  
An obvious candidate for an experimental implementation of this class of decoupling operators is a Hamiltonian 
$\hat{H}_p = \hbar \chi \hat{b}^\dag \hat{b}$ with a 
parameter $\chi$ that is activated for a time $t_p$ such that $\chi t_p = \varphi$. Then the induced unitary evolution operator is 
\begin{equation}
U(t_p) = e^{-i\chi \hat{b}^\dagger \hat{b} t_p } = \hat{p}
\end{equation}
 as required. Note that the Hamiltonian of the harmonic oscillator contains a term of exactly this nature: 
$\hbar \omega_t \hat{b}^\dagger \hat{b}$. Unfortunately this term does not commute with the rest of the interaction 
Hamiltonian and therefore does not act undisturbed, otherwise it would implement a perfect decoupling pulse on its own. 
Even so, the presence of this term does imply a sort of self-decoupling that depends on the trap frequency $\omega_t$: 
for very high frequencies the term $\hbar \omega_t \hat{b}^\dagger \hat{b}$ dominates the Hamiltonian and can thus implement the 
decoupling pulse almost perfectly. However, with decreasing frequency the interacting parts of the Hamiltonian disturb 
the purity of the pulse. This offers
 another view on why a higher trap frequency improves the overall fidelity of the entanglement process.

Still, for lower trap frequencies $\omega_t$ this gives us an idea of how to implement the Hamiltonian $\hat{H}_p$: 
In our scenario a possibility is to switch off interactions during short time intervals of motion $t_p$ during 
the interaction time $\tau$, such that within the time interval $t_p$ only the term $\hbar \omega_t \hat{b}^\dag \hat{b}$ remains in the 
interaction picture. In the Lamb-Dicke 
regime this could be achieved by a Stark-switching procedure, since the coupling of the internal states with the centre-of-mass motion 
without
a radiation field 
is small during the interaction time $\tau$. This has the additional effect that the time used to implement the pulses does not 
contribute to the interaction time $\tau$, since no interaction is taking place. Therefore, the whole process now takes a time 
$T_p = \tau + N t_p$ depending on the number of pulses $N$. Keep in mind, though, that the time $T_p$ cannot grow arbitrarily large 
due to experimental constraints. When $T_p$ grows larger, spontaneous emission will eventually become a problem. Therefore, there 
is a practical limit on the time $N t_p$ available to implement all of the pulses. If $\Gamma$ is the rate of spontaneous decay of 
the internal state $\ket{2}$ of either material qubit, 
then we require that 
\begin{equation}
T_p = \tau + N t_p \ll \frac{1}{\Gamma} .
\end{equation}
Since the interaction time $\tau$ is determined by the occurence of the collapse phenomenon and is of the order $\tau \sim \frac{1}{2 |g|}$,
 we can roughly estimate that the available time to implement our decoupling pulses is limited by
\begin{equation} \label{limited}
N t_p \ll \frac{|g|}{\Gamma} \tau ,
\end{equation}
where $\frac{|g|}{\Gamma}$ depends on the specific experimental setup. Recent experimental developments look very promising: 
whereas in $2003$ an experiment by McKeever {\it et al.} \cite{McKeever} achieved the ratio 
$\frac{|g|}{\Gamma}=6.15$, in $2007$ an experiment by Colombe {\it et al.} \cite{Colombe} was performed with a significantly 
improved ratio of $\frac{|g|}{\Gamma}=71.66$.

This leads to the question of how large $N$ and $t_p$ need to be to see a positive effect of the decoupling procedure.
Remember that the class of operators $\hat{p}$ was derived in the continuous control limit where $N \to \infty$ 
and $\Delta t \to 0$. As a consequence, very high repetitions of applications of $\hat{p}$ may be necessary to observe a 
positive effect of the decoupling procedure. In order to examine just how large $N$ should be and what phase $\varphi$ is preferable 
for the decoupling operator $\hat{p}$, we will look at some numerical simulations in the next section.

\subsection{Numerical simulation}

\begin{figure}[b]
\begin{center}
\includegraphics[width=7cm]{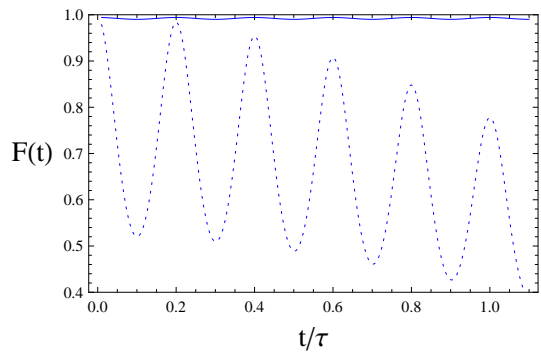}
\caption{\label{fig-4} The fidelity $\mathcal{F}(t)$ of the system compared to its ideal
evolution during the interaction time $\tau$. Without decoupling (dotted line) there is a steady drop
in the fidelity and also oscillations with frequency $2 \omega_t$ as explained by the induced
oscillations in the coherent motion states. With active parity kick decoupling, using 200 $\hat{\mathcal{P}}$  
pulses in total during the interaction time, the system effectively remains at unit fidelity
throughout the process.}
\end{center}
\end{figure}

We have run a numerical simulation for the trapped system under the 
influence of the interaction Hamiltonian $\hat{H}_I$.
For our simulation we assume that the material qubit and the radiation field are initially in the state 
\begin{equation}
\ket{\Psi_0} = \frac{1}{\sqrt{2}} ( \ket{0} \ket{\alpha}_c\ket{0}_t + \ket{1} \ket{\alpha}_c\ket{0}_t ),
\end{equation}
meaning that the centre-of-mass motion is in the oscillator ground state while the internal states are in the superposition 
$\frac{1}{\sqrt{2}} (\ket{0} + \ket{1})$ and the driving field is in the coherent state $\ket{\alpha}_c$ with $|\alpha|^2 = 100$ the 
mean photon number. The coupling strengths are chosen such that $\frac{\vert \gamma \vert}{\vert g \vert} = 0.4$ 
and $\frac{\omega_t}{|g|} = 10$. Figure \ref{fig-4} shows plots of the fidelity 
\begin{equation}
\mathcal{F}(t) = \left\vert \bra{\Psi_0} \hat{U}_\text{id}^\dag(t) \hat{U}(t) \ket{\Psi_0} \right\vert^2
\end{equation}
over the course of the interaction time $\tau$,
comparing the time evolution under the actual Hamiltonian $\hat{H}_I$ and the ideal Hamiltonian $\hat{H}_\text{id}$. 
First is plotted the fidelity as it evolves without decoupling. There is some oscillation with a frequency of 
$2\omega_t$, and one can clearly see that the fidelity is steadily decreasing. The oscillation is expected due to the oscillatory behaviour 
in the coherent state displacement, see \eref{displacement}. The second plot 
demonstrates the effect of our decoupling scheme, where we chose the parity operator $\hat{\mathcal{P}}$ as the decoupling operator and 
applied it evenly 200 times over 
the whole interaction time. There is no visible drop of the fidelity, and even the minimal points of the still present oscillation 
are well above $\mathcal{F} > 0.99$.

\begin{figure}[b]
\begin{center}
\includegraphics[width=7cm]{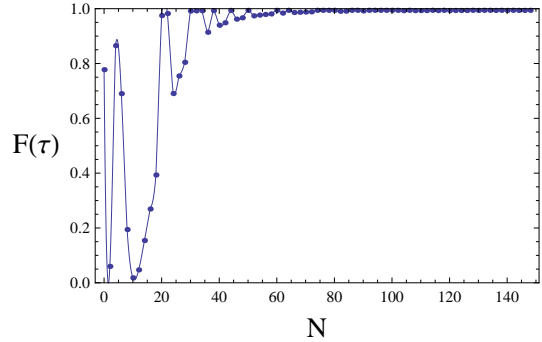}
\caption{\label{fig-5} The final fidelity $\mathcal{F}(\tau)$ of the system at the end of the interaction depending on the number $N$ of 
parity kicks $\hat{\mathcal{P}}$ used. The fidelity stabilizes at 
$N \sim 50$ kicks at high fidelity values.}
\end{center}
\end{figure}
\begin{figure}[tb]
\begin{center}
\includegraphics[width=7cm]{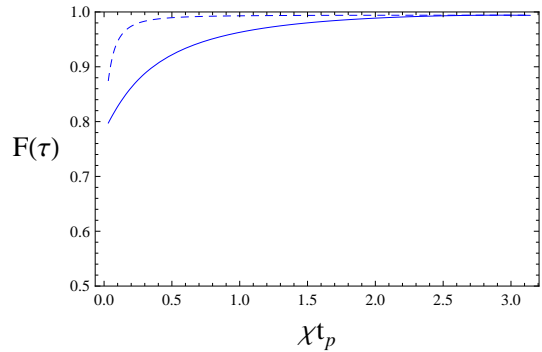}
\caption{\label{fig-6} The final fidelity $\mathcal{F}(\tau)$ of the system at the end of the interaction depending on the parameter $\chi$ 
when using a Hamiltonian $\hat{H}_p = \hbar \chi \hat{b}^\dag \hat{b}$ to implement a non-parity decoupling pulse $\hat{p}$. 
The fidelity $\mathcal{F}(\tau)$ is plotted for different numbers of pulses $N=50$ (solid) and $N=400$ (dotted line). It improves with 
$\chi t_p$ approaching the parity operator value $\pi$. The fidelity is also generally higher for higher number of pulses used.}
\end{center}
\end{figure}
Encouraged by this result we studied how many parity kicks one actually needs to achieve a high fidelity at the end of the interaction. 
We therefore ran additional simulations calculating the final fidelity $\mathcal{F}(\tau)$ depending on the number $N$ of parity kicks 
employed during that time. 
The results are plotted in figure \ref{fig-5}; as one can see the fidelity stabilizes on a high level at around $N \sim 50$ parity kicks. 
Below that threshold the fidelity is unpredictable which suggests that the time between pulses
$\Delta t$ is too high and, as a consequence, the system evolution is governed by higher terms
of the average Hamiltonian $\hat{H}_N$.

But the parity operator is only one special case of the class of decoupling operators we found. Indeed, in the experimental realization 
we proposed the parity operator might need an unacceptably long time $t_p$ to be implemented. Therefore, we ran additional simulations 
with a decoupling pulse implemented the Hamiltonian $\hat{H}_p = \hbar \chi \hat{b}^\dag \hat{b}$ over a time $t_p$, as explained previously. 
We plotted the dependency of the fidelity $\mathcal{F}(\tau)$ after the interaction time $\tau$ on the phase $\chi t_p$ for different 
numbers $N$ of total pulses. 
The results are shown in figure \ref{fig-6}. As we can see, the fidelity improves the closer $\chi t_p$ comes to the parity value 
$\pi$, which makes the parity operator $\hat{\mathcal{P}}$ the preferred choice for the decoupling procedure. The fidelity also improves with
the number of pulses $N$, so the smaller the parameter $\chi t_p$ is in the experimental setup, the more pulses must be employed to get a 
good fidelity at the end of the interaction.

\begin{figure}[b]
\begin{center}
\includegraphics[width=7cm]{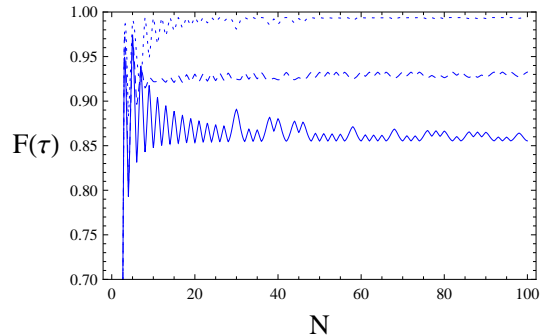}
\caption{\label{fig-7} The final fidelity $\mathcal{F}(\tau)$ of the system at the end of the interaction with time-constrained decoupling.
The fidelity is plotted over the number of pulses used, while the number of pulses $N$ also determines the pulse width $\omega_t t_p$ - the 
higher $N$, the shorter $t_p$. The solid curve shows the final fidelity assuming that the additional time $N t_p$ to implement pulses is 
$\tau$, whereas the dashed curve assumes it to be $2\tau$ and the dotted curve is for $4 \tau$. All curves converge to an almost constant 
fidelity value at higher pulse numbers, but show unpredictable behaviour below $N=20$ pulses. The more time is available for pulse 
implementation, the higher the achieved fidelity.}
\end{center}
\end{figure}

But as explained before, in actual experimental realizations the number of pulses one can implement is not independent of the pulse width 
$\chi t_p$ due to constraints on the overall process time $T_p$, expected to be primarily given by the rate of spontaneous decay $\Gamma$.
 Given this constraint,  we need to figure out what the best choice of number of pulses $N$ is, considering that the choice of $N$ also fixes the maximal pulse 
time $t_p$ by the inequality in \eref{limited}. We ran simulations under the assumptions that the overall process time $T_p=\tau+Nt_p$ 
is $2\tau$, $3\tau$ and $5\tau$, respectively. The results are shown in figure \ref{fig-7}. Unsurprisingly the results are 
better if more time is available for pulse implementation. Somewhat surprsingly, however, is that the achievable fidelity stabilizes at
 higher pulse numbers $N$, so the choice of whether to do larger numbers of short pulses or smaller numbers of longer pulses has little 
influence as long as the number of pulses does not fall below a certain threshold. For small numbers of pulses the results are unpredictable,
 suggesting that the delay between pulses $\Delta t$ is large enough that higher orders of the average Hamiltonian govern the time evolution.
 For optimal results, judging from our combined numerical simulations, we recommend to aim for $N=50$ pulses and then make the pulses as 
close to the parity operator as possible.

\subsection{Improving fidelity for lower number of pulses}
In the case where even $N=50$ pulses present a technical challenge in an experimental realization, there may be a way to decrease 
the necessary number of pulses further. In 2007 a novel decoupling sequence was presented by Uhrig \cite{Uhrig} for protecting a single
 quantum bit with a sequence of $\pi$ pulses which are not applied equidistantly over time with a common pulse distance of $\Delta t$. 
Instead, the pulses are applied at times $t_j = \tau \cdot \sin^2\left[ \pi j / (2N+2) \right]$ during the interaction time $\tau$. 
These times are derived such that they eliminate higher orders of $\hat{H}_N$, whereas so far we only considered the lowest order 
$\hat{\mathcal{H}}$ 
which remains in the limit $N \to \infty$. Although Uhrig derived these specific times for a specific scenario with $\pi$ pulses on a
 single qubit, we confirmed in numerical simulations that they provide an improvement in our case, as well, particularly when using 
the parity kick.

\begin{figure}
\begin{center}
\includegraphics[width=7cm]{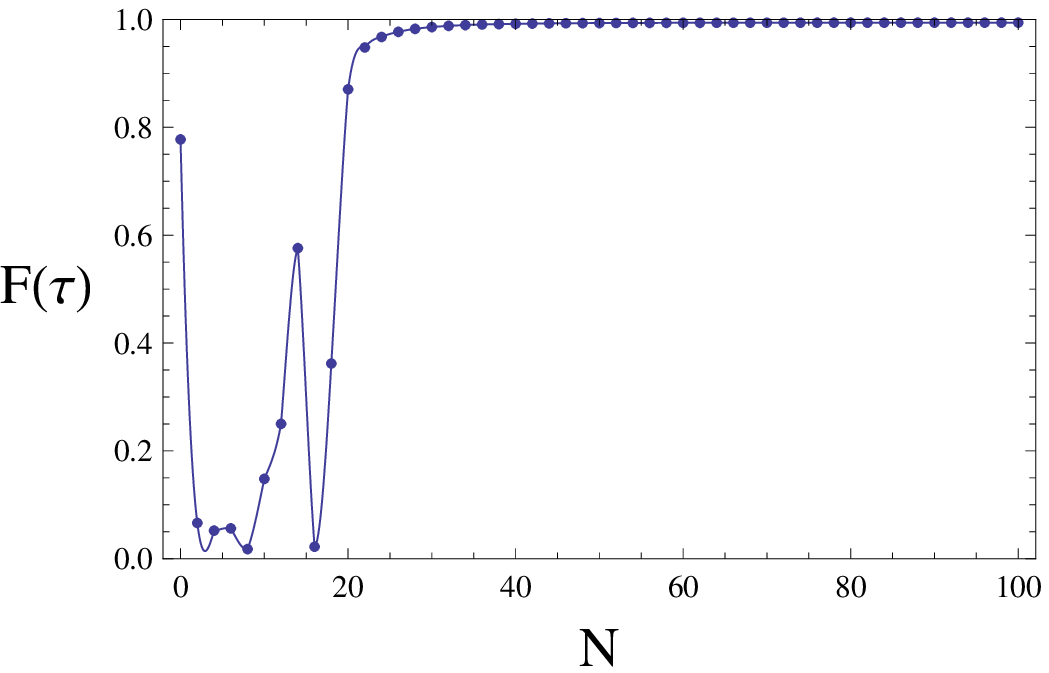}
\includegraphics[width=7cm]{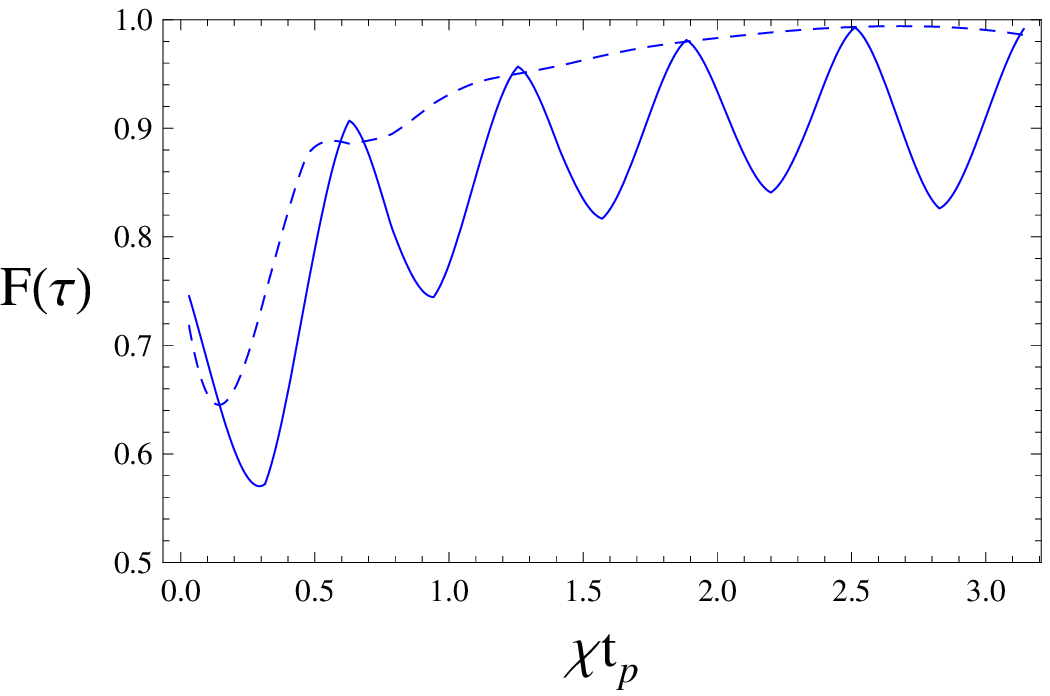}
\caption{\label{fig-8} The final fidelity $\mathcal{F}(\tau)$ when employing Uhrig's dynamical decoupling. 
Figure a) shows final fidelity depending on the number $N$ of parity kicks employed (compare also with figure \ref{fig-5}), while figure b) compares equidistant decoupling (solid line) with Uhrig decoupling (dashed line) for $N=30$ general decoupling pulses depending on the parameter $\chi t_p$. Uhrig decoupling shows improved stability for lower number of pulses compared to equidistant decoupling.}
\end{center}
\end{figure}

Figure \ref{fig-8}a shows the final fidelity achievable when using parity kicks with the Uhrig decoupling sequence. Compared to 
figure \ref{fig-5} we see that Uhrig's decoupling shows improvements in achieved fidelity and stability particularly in the range 
between $N=20$ and $N=50$ parity kicks. Inspired by these findings, we also did another numerical run for the situation of figure 6 
where we investigated the effectiveness of our more general class of decoupling pulses, but this time with only $N=30$ pulses. In figure 
\ref{fig-8}b the results are shown depending on the parameter $\chi t_p$, comparing our standard, equidistant decoupling with the Uhrig 
sequence. As expected, the standard method shows signs of instability at low numbers of pulses, making the achievable final fidelity 
hard to predict. This hints at higher orders of $\hat{H}_N$ dominating the time evolution. Here the Uhrig sequence works better as it 
was designed 
to eliminate more of those higher orders. The result is a more stable and predictable curve for these low number of pulses. However, 
with $N \ge 50$ we found no further 
advantage from employing Uhrig's sequence, so its advantages are strictly limited to scenarios where only a small number of pulses can be 
implemented.

\section{Conclusions}

In this paper we considered the influence of the centre-of-mass motion of material qubits (modelled by three-level systems) in 
an implementation of a hybrid quantum repeater. 
This motion is a source of decoherence and dissipation and decreases the probabilities of creating high fidelity entangled pairs 
of distant qubits.   

In particular, we investigated two distant three-level systems confined in harmonic potentials and coupled to
single-mode cavity fields, which are connected by a multi-mode 
optical fibre. For the description of the centre-of-mass motion of the three-level systems we used the Lamb-Dicke and the rotating wave 
approximations for the description of the interaction between the radiation field and the internal states. We further simplified by setting 
the single-mode radiation field frequencies to be equal to the frequency difference of the two upper internal energy levels of the 
three-level system. In this model we calculated the time evolution of a Ramsey-type coupling sequence and we determined 
the optimal POVM measurements which have 
to be performed on the radiation field in order to postselect a Bell pair with minimal error. 
Within this model we found that general effects of the centre-of-mass motion of the qubits lower
success probabilities and achievable fidelities significantly. Nevertheless, these quantities are
very sensitive to changes of the harmonic potential frequencies involved. A small increase in the frequencies can improve the probabilities
of creating high fidelity entangled pairs. In the case of high trap frequencies our results resemble almost the ideal case of motionless 
qubits, for which during the 
collapse phenomenon  a perfect Bell state can be created with $25\%$ probability \cite{Bernad}. Thus high trap frequencies suppress 
effects of the qubits 
centre-of-mass motion.
    
If sufficiently high trap frequencies are not achievable, a suppression of the decohering effects of the qubits' centre-of-mass
motion can be achieved by dynamical decoupling methods. For this purpose we derived a class of
appropiate unitary control operations, which also contain the well known parity kick operation as a special case. This dynamical decoupling 
may be achieved by interrupting the matter-field interactions inside the photonic cavities for short time intervals with the help of
Stark switching techniques, for example. Our simulations demonstrate that approximately $50$ such interruptions during the 
matter-field interaction time are satisfactory to achive a sufficient degree of suppression of the decohering influence of the effects
of the qubits' centre-of-mass motion. Even lower numbers can be made to work if one switches to the Uhrig decoupling sequence.

\ack

This work is supported by the BMBF project QK{\_}QuOReP. 

\appendix
 \section{Baker-Hausdorff formulas}
 \label{BH}
In this appendix we derive a general Baker-Hausdorff identity which is used in equation \eref{evolution} in the main text.
Let us define the unitary operator
\begin{equation}
 \hat{U}(t)=\ee^{i \omega_t \hat{b}^\dagger \hat{b} t} \ee^{-i \omega_t \hat{b}^\dagger \hat{b} t-i |\gamma| \sqrt{n} 
\big(\hat{b}+\hat{b}^\dagger\big) t},
\end{equation}
which fulfills the following equation of motion
\begin{eqnarray}
\label{Ueqmotion}
 \frac{d \hat{U}(t)}{dt}&=&-i |\gamma| \sqrt{n} \ee^{i \omega_t \hat{b}^\dagger \hat{b} t} \big(\hat{b}+\hat{b}^\dagger\big) 
\ee^{-i \omega_t \hat{b}^\dagger \hat{b} t} \hat{U}(t) \nonumber \\
&=&\Big(\hat{A}(t)+\hat{B}(t)\Big)\hat{U}(t),
\end{eqnarray}
where
\begin{eqnarray}
 \hat{A}(t)&=& -i |\gamma| \sqrt{n} \ee^{-i \omega_t t} \hat{b}, \\
 \hat{B}(t)&=& -i |\gamma| \sqrt{n} \ee^{i \omega_t t} \hat{b}^\dagger.
\end{eqnarray}
These operators obey the commutation relations
\begin{eqnarray}
 \left[\hat{A}(t),\hat{A}(t')\right]&=&\left[\hat{B}(t),\hat{B}(t')\right]=0, \\
 \left[\hat{A}(t),\hat{B}(t')\right]&=&-|\gamma|^2 n \ee^{-i \omega (t-t')}.
\end{eqnarray}
Consider now the operator $\hat{V}$ defined as
\begin{equation}
 \hat{V}(t)=\ee^{\int^t_0 dt' \hat{B}(t')}\ee^{\int^t_0 dt' \hat{A}(t')},
\end{equation}
whose equation of motion is
\begin{eqnarray}
 &&\frac{d \hat{V}(t)}{dt}=\ee^{\int^t_0 dt' \hat{B}(t')}\Big(\hat{B}(t)+\hat{A}(t)\Big)\ee^{\int^t_0 dt' \hat{A}(t')} \nonumber \\
&&=\Big(\hat{B}(t)+\ee^{\int^t_0 dt' \hat{B}(t')}\hat{A}(t)\ee^{-\int^t_0 dt' \hat{B}(t')} \Big)\hat{V}.
\end{eqnarray}
Now using the identity
\begin{equation}
 \ee^{\hat{B}}\hat{A}\ee^{-\hat{B}}=\hat{A}+\left[\hat{B},\hat{A}\right]+\frac{1}{2!} \left[\hat{B},\left[\hat{B},\hat{A}\right] \right] 
+ \dots
\end{equation}
and the fact that the commutator of $\hat{A}(t)$ and $\hat{B}(t)$ is a number, the equation of motion for $\hat{V}$ takes the form
\begin{equation}
\label{Veqmotion}
 \frac{d \hat{V}(t)}{dt}=\Big(\hat{B}(t)+\hat{A}(t)+\int^t_0 dt'\left[\hat{B}(t'),\hat{A}(t)\right] \Big)\hat{V}.
\end{equation}
Now comparing \eref{Ueqmotion} with \eref{Veqmotion} and with the aid that $\left[\hat{B}(t'),\hat{A}(t)\right]$ is a number, we get
\begin{equation}
 \hat{U}(t)=\ee^{\int^t_0 dt' \hat{B}(t')}\ee^{\int^t_0 dt' \hat{A}(t')} \ee^{-\int^{t}_o dt' 
\int^{t'}_0 dt''\left[\hat{B}(t''),\hat{A}(t')\right]}.
\end{equation}
Another useful formula can be obtained
\begin{eqnarray}
&&\ee^{-i \omega_t \hat{b}^\dagger \hat{b} t}\ee^{\int^t_0 dt' \hat{B}(t')}
\ee^{i \omega_t \hat{b}^\dagger \hat{b} t}
\ee^{-i \omega_t \hat{b}^\dagger \hat{b} t}\ee^{\int^t_0 dt' \hat{A}(t')}\ee^{i \omega_t \hat{b}^\dagger \hat{b} t} \ee^{-\int^{t}_o dt' 
\int^{t'}_0 dt''\left[\hat{B}(t''),\hat{A}(t')\right]} \nonumber \\
=&& 
\ee^{-i |\gamma| \sqrt{n}\int^t_0 dt' \ee^{i \omega_t t'} \ee^{-i \omega_t t} \hat{b}^\dagger } 
\ee^{-i |\gamma| \sqrt{n}\int^t_0 dt' \ee^{-i \omega_t t'} \ee^{i \omega_t t} \hat{b} }
\ee^{-|\gamma|^2 n \int^{t}_o dt' 
\int^{t'}_0 dt''\ee^{i \omega (t''-t')}} \nonumber \\
=&& \ee^{-\frac{|\gamma|\sqrt{n}}{\omega_t}\left(1-\ee^{-i\omega_t t}\right) \hat{b}^\dagger}
\ee^{\frac{|\gamma|\sqrt{n}}{\omega_t}\left(1-\ee^{i\omega_t t}\right) \hat{b}}
\ee^{-\frac{|\gamma|^2n}{\omega^2_t}\left(1-i\omega_t t -\ee^{-i\omega_t t} \right)} \nonumber \\
=&&\ee^{i\Phi_n(t)} \ee^{-\alpha_n(t)\hat{b}^\dagger+\alpha^*_n(t)\hat{b}}= 
\ee^{-i \omega_t \hat{b}^\dagger \hat{b} t} \hat{U}(t)\ee^{i \omega_t \hat{b}^\dagger \hat{b} t},
\end{eqnarray}
where we used the identity
\begin{equation}
\ee^{\ee^{-i\hat{B}}\hat{A}\ee^{i\hat{B}}}=\sum^\infty_{n=0}\frac{\left(\ee^{-i\hat{B}}\hat{A}\ee^{i\hat{B}}\right)^n}{n!} 
=\sum^\infty_{n=0}\ee^{-i\hat{B}}\frac{\hat{A}^n}{n!}\ee^{i\hat{B}}=\ee^{-i\hat{B}}\ee^{\hat{A}}\ee^{i\hat{B}},
\end{equation}
and introduced the parameters
\begin{eqnarray}
 \alpha_n(t)&=&\frac{|\gamma|\sqrt{n}}{\omega_t}\left(1-\ee^{-i\omega_t t}\right), \\
 \Phi_n(t)&=&\frac{|\gamma|^2n}{\omega^2_t}\left(\omega_t t-\sin(\omega_t t)\right).
\end{eqnarray}

\section{The states of the radiation field emerging from the interactions}
\label{Calc}
In this appendix we present the detailed structure of the states $\ket{g_{ij}(t)}$ ($i,j=0,1,2$) which appear in equation \eref{Psi} in the main text. They are defined by:
\begin{eqnarray}
\ket{g_{10}(t)}&=&\sum^{\infty}_{n=0}
\Big(\frac{1}{\sqrt{2}}g^+_1(n,\tau)\ket{\alpha_n(\tau)\ee^{-i \omega_t (T+\tau)}}_{t,A}+\frac{1}{\sqrt{2}}g^-_1(n,\tau)
\ket{-\alpha_n(\tau)\ee^{-i \omega_t (T+\tau)}}_{t,A}
 \Big) \nonumber \\
&\times&\ket{n}_{c,B} \ket{0}_{t,B}\ee^{-i\omega_c n t}, \nonumber \\
\ket{g_{01}(t)}&=&\sum^{\infty}_{n=0} \Big(\frac{1}{\sqrt{2}}g^+_1(n,\tau)\ket{\alpha_n(\tau)}_{t,B}+\frac{1}{\sqrt{2}}g^-_1(n,\tau)
\ket{-\alpha_n(\tau)}_{t,B}
 \Big) \ket{n}_{c,B} \ket{0}_{t,A}\ee^{-i\omega_c n t},\nonumber  
\end{eqnarray}
\begin{eqnarray}
\ket{g_{20}(t)}&=& \sum^{\infty}_{n=0}\Big(\frac{1}{\sqrt{2}}g^+_2(n,\tau)\ket{\alpha_{n+1}(\tau)\ee^{-i \omega_t (T+\tau)}}_{t,A}
\nonumber \\
&+&\frac{1}{\sqrt{2}}g^-_2(n,\tau)
\ket{-\alpha_{n+1}(\tau)\ee^{-i \omega_t (T+\tau)}}_{t,A}
 \Big)  \ket{n}_{c,B} \ket{0}_{t,B}\ee^{-i\omega_c n t}, \nonumber
\\ 
\ket{g_{02}(t)}&=&\sum^{\infty}_{n=0} \Big(\frac{1}{\sqrt{2}}g^+_2(n,\tau)\ket{\alpha_{n+1}(\tau)}_{t,B}+\frac{1}{\sqrt{2}}g^-_2(n,\tau)
\ket{-\alpha_{n+1}(\tau)}_{t,B}
 \Big)\nonumber \\
&\times& \ket{n}_{c,B} \ket{0}_{t,A}\ee^{-i\omega_c n t}, \nonumber
\end{eqnarray}
\begin{eqnarray}
\ket{g_{11}(t)}&=&\sum^{\infty}_{n=0}\Big(\frac{g^+_1(n,\tau)}{2\sqrt{2}}\ee^{i|g|\sqrt{n} \tau+\Phi_n(\tau)}\ket{\alpha_n(\tau)
\ee^{-i \omega_t (T+\tau)}}_{t,A} \ket{\alpha_n(\tau)}_{t,B} \nonumber \\ 
&+&\frac{g^+_1(n,\tau)}{2\sqrt{2}}\ee^{-i|g|\sqrt{n} \tau+\Phi_n(\tau)} \ket{\alpha_n(\tau)\ee^{-i \omega_t (T+\tau)}}_{t,A} 
\ket{-\alpha_n(\tau)}_{t,B} \nonumber \\
&+&\frac{g^-_1(n,\tau)}{2\sqrt{2}}\ee^{i|g|\sqrt{n} \tau+\Phi_n(\tau)}\ket{-\alpha_n(\tau)\ee^{-i \omega_t (T+\tau)}}_{t,A} 
\ket{\alpha_n(\tau)}_{t,B} \nonumber \\ 
&+&\frac{g^-_1(n,\tau)}{2\sqrt{2}}\ee^{-i|g|\sqrt{n} \tau+\Phi_n(\tau)}\ket{-\alpha_n(\tau)\ee^{-i \omega_t (T+\tau)}}_{t,A} 
\ket{-\alpha_n(\tau)}_{t,B} \Big) \ket{n}_{c,B}\ee^{-i\omega_c n t}, \nonumber 
\end{eqnarray}
\begin{eqnarray}
\ket{g_{12}(t)}&=&\sum^{\infty}_{n=0} \Big(\frac{-g^+_1(n+1,\tau)}{2\sqrt{2}}\ee^{i|g|\sqrt{n+1}\tau+\Phi_{n+1}(\tau)}\ket{\alpha_{n+1}(\tau)
\ee^{-i \omega_t (T+\tau)}}_{t,A} 
\ket{\alpha_{n+1}(\tau)}_{t,B} \nonumber \\
&+&\frac{g^+_1(n+1,\tau)}{2\sqrt{2}}\ee^{-i|g|\sqrt{n+1}\tau+\Phi_{n+1}(\tau)} \ket{\alpha_{n+1}(\tau)\ee^{-i \omega_t (T+\tau)}}_{t,A} 
\ket{-\alpha_{n+1}(\tau)}_{t,B} \nonumber \\ 
&-&\frac{g^-_1(n+1,\tau)}{2\sqrt{2}}\ee^{i|g|\sqrt{n+1}\tau+\Phi_{n+1}(\tau)}\ket{-\alpha_{n+1}(\tau)\ee^{-i \omega_t (T+\tau)}}_{t,A}
 \ket{\alpha_{n+1}(\tau)}_{t,B} \nonumber \\ 
&+&\frac{g^-_1(n+1,\tau)}{2\sqrt{2}}\ee^{-i|g|\sqrt{n+1}\tau+\Phi_{n+1}(\tau)}\ket{-\alpha_{n+1}(\tau)\ee^{-i \omega_t (T+\tau)}}_{t,A} 
\ket{-\alpha_{n+1}(\tau)}_{t,B} \Big)\nonumber \\
&\times& \ket{n}_{c,B}\ee^{-i\omega_c n t}, \nonumber
\end{eqnarray}
\begin{eqnarray}
\ket{g_{21}(t)}&=&\sum^{\infty}_{n=0}\Big(\frac{g^+_2(n,\tau)}{2\sqrt{2}}\ee^{i|g|\sqrt{n}\tau+\Phi_n(\tau)}\ket{\alpha_{n+1}(\tau)
\ee^{-i \omega_t (T+\tau)}}_{t,A} \ket{\alpha_n(\tau)}_{t,B} \nonumber \\
&+&\frac{g^+_2(n,\tau)}{2\sqrt{2}}\ee^{-i|g|\sqrt{n}\tau+\Phi_n(\tau)} \ket{\alpha_{n+1}(\tau)\ee^{-i \omega_t (T+\tau)}}_{t,A} 
\ket{-\alpha_n(\tau)}_{t,B} \nonumber \\
&+& \frac{g^-_2(n,\tau)}{2\sqrt{2}}\ee^{i|g|\sqrt{n}\tau+\Phi_n(\tau)}\ket{-\alpha_{n+1}(\tau)\ee^{-i \omega_t (T+\tau)}}_{t,A} 
\ket{\alpha_n(\tau)}_{t,B} \nonumber \\ 
&+&\frac{g^-_2(n,\tau)}{2\sqrt{2}}\ee^{-i|g|\sqrt{n}\tau+\Phi_n(\tau)}\ket{-\alpha_{n+1}(\tau)\ee^{-i \omega_t (T+\tau)}}_{t,A} 
\ket{-\alpha_n(\tau)}_{t,B} \Big) \nonumber \\
&\times&\ket{n}_{c,B}\ee^{-i\omega_c n t}, \nonumber
\end{eqnarray}
\begin{eqnarray}
\ket{g_{22}(t)}&=&\sum^{\infty}_{n=0} \Big(\frac{-g^+_2(n+1,\tau)}{2\sqrt{2}}\ee^{i|g|\sqrt{n+1}\tau+\Phi_{n+1}(\tau)}\ket{\alpha_{n+2}(\tau)
\ee^{-i \omega_t (T+\tau)}}_{t,A} \ket{\alpha_{n+1}(\tau)}_{t,B} \nonumber \\ 
&+&\frac{g^+_2(n+1,\tau)}{2\sqrt{2}}\ee^{-i|g|\sqrt{n+1}\tau+\Phi_{n+1}(\tau)} \ket{\alpha_{n+2}(\tau)\ee^{-i \omega_t (T+\tau)}}_{t,A} 
\ket{-\alpha_{n+1}(\tau)}_{t,B} \nonumber \\
&-&\frac{g^-_2(n+1,\tau)}{2\sqrt{2}}\ee^{i|g|\sqrt{n+1}\tau+\Phi_{n+1}(\tau)}\ket{-\alpha_{n+2}(\tau)\ee^{-i \omega_t (T+\tau)}}_{t,A}
 \ket{\alpha_{n+1}(\tau)}_{t,B} \nonumber \\ 
&+&\frac{g^-_2(n+1,\tau)}{2\sqrt{2}}\ee^{-i|g|\sqrt{n+1}\tau+\Phi_{n+1}(\tau)}\ket{-\alpha_{n+2}(\tau)\ee^{-i \omega_t (T+\tau)}}_{t,A} 
\ket{-\alpha_{n+1}(\tau)}_{t,B} \Big) \nonumber \\
&\times&\ket{n}_{c,B}\ee^{-i\omega_c n t}, \nonumber
\end{eqnarray}
with the normalized photon number states $\ket{n}_c$ ($n\in {\mathbb N}$) and
\begin{eqnarray}
 && g^+_1(n,t)= f_n
\ee^{i \frac{|\gamma|^2n}{\omega^2_t}\left(\omega_t t-\sin(\omega_t t)\right)} \ee^{i|g|\sqrt{n}t},  \nonumber \\
&& g^-_1(n,t)= f_n 
\ee^{i \frac{|\gamma|^2n}{\omega^2_t}\left(\omega_t t-\sin(\omega_t t)\right)} \ee^{-i|g|\sqrt{n}t},  \nonumber \\
&& g^+_2(n,t)=-f_{n+1}\frac{g}
{|g|} 
\ee^{i \frac{|\gamma|^2(n+1)}{\omega^2_t}\left(\omega_t t-\sin(\omega_t t)\right)}\ee^{i|g|\sqrt{n+1}t},  \nonumber \\
&& g^-_2(n,t)=f_{n+1}\frac{g}
{|g|} 
\ee^{i \frac{|\gamma|^2(n+1)}{\omega^2_t}\left(\omega_t t-\sin(\omega_t t)\right)}\ee^{-i|g|\sqrt{n+1}t}, \nonumber \\
&&f_n=\ee^{-|\alpha|^2/2}\frac{\alpha^n}{\sqrt{n!}} \frac{1}{2 \sqrt{2}}. \nonumber
\end{eqnarray}
We are going to use the following identities
\begin{eqnarray}
 \mathrm{Tr}\{\ket{\alpha_n(\tau)}\bra{\alpha_m(\tau)}\}&=&\ee^{-\frac{|\gamma|^2}{\omega^2_t}
\Big(1-\cos(\omega_t \tau)\Big)\big(\sqrt{n}-\sqrt{m}\big)^2}, \nonumber \\
\mathrm{Tr}\{\ket{-\alpha_n(\tau)}\bra{\alpha_m(\tau)}\}&=&
\mathrm{Tr}\{\ket{\alpha_n(\tau)}\bra{-\alpha_m(\tau)}\} \nonumber \\
&=&\ee^{-\frac{|\gamma|^2}{\omega^2_t}
\Big(1-\cos(\omega_t \tau)\Big)\big(\sqrt{n}+\sqrt{m}\big)^2}. \nonumber
\end{eqnarray}
The coefficients of the field states in \eref{fieldstate}, $ \hat{\rho}_F(t) 
 = \sum^2_{i,j=0} \sum^{\infty}_{n,m=0} a_{ij}(n,m)\ket{n}\bra{m}$,
emerged from the three-step interaction are defined by
\begin{eqnarray}
 &&a_{00}(n,m)=\frac{1}{4}\ee^{-|\alpha|^2}\frac{\alpha^{n}(\alpha^*)^{m}}{\sqrt{n!}\sqrt{m!}} \ee^{-i\omega_c t (n-m)}, \nonumber \\
 &&a_{10}(n,m)=a_{01}(n,m)= \frac{1}{8} \ee^{-|\alpha|^2} \frac{\alpha^n \big(\alpha^* \big)^m}{\sqrt{n!}\sqrt{m!}}\times \nonumber \\
&&\times\Big[\cos\big(|g|(\sqrt{n}-\sqrt{m})\tau\big)\ee^{-\frac{|\gamma|^2}{\omega^2_t}
\Big(1-\cos(\omega_t \tau)\Big)\big(\sqrt{n}-\sqrt{m}\big)^2}+\cos\big(|g|(\sqrt{n}+\sqrt{m})\tau\big)\times \nonumber \\
&&\times\ee^{-\frac{|\gamma|^2}{\omega^2_t}
\Big(1-\cos(\omega_t \tau)\Big)\big(\sqrt{n}+\sqrt{m}\big)^2} \Big]\ee^{-i\Big(\omega_c t-
\frac{|\gamma|^2}{\omega^2_t}\left(\omega_t \tau-\sin(\omega_t \tau)\right)\Big)(n-m)},\nonumber 
\end{eqnarray}
\begin{eqnarray}
 &&a_{20}(n,m)=a_{02}(n,m)=\frac{1}{8}\ee^{-|\alpha|^2}\frac{\alpha^{n+1}(\alpha^*)^{m+1}}{\sqrt{(n+1)!}\sqrt{(m+1)!}} \times \nonumber \\
&&\times \Big[\cos\big(|g|(\sqrt{n+1}-\sqrt{m+1})\tau\big)\ee^{-\frac{|\gamma|^2}{\omega^2_t}
\Big(1-\cos(\omega_t \tau)\Big)\big(\sqrt{n+1}-\sqrt{m+1}\big)^2} \nonumber \\
&&-\cos\big(|g|(\sqrt{n+1}+\sqrt{m+1})\tau\big)\ee^{-\frac{|\gamma|^2}{\omega^2_t}
\Big(1-\cos(\omega_t \tau)\Big)\big(\sqrt{n+1}+\sqrt{m+1}\big)^2} \Big] \times \nonumber \\
&&\times \ee^{-i\Big(\omega_c t-
\frac{|\gamma|^2}{\omega^2_t}\left(\omega_t \tau-\sin(\omega_t \tau)\right)\Big)(n-m)}, \nonumber 
\end{eqnarray}
\begin{eqnarray}
 &&a_{11}(n,m)=\frac{1}{16} \ee^{-|\alpha|^2} \frac{\alpha^n \big(\alpha^* \big)^m}{\sqrt{n!}\sqrt{m!}}
\Big[\cos^2\big(|g|(\sqrt{n}-\sqrt{m})\tau\big)\ee^{-\frac{2|\gamma|^2}{\omega^2_t}
\Big(1-\cos(\omega_t \tau)\Big)\big(\sqrt{n}-\sqrt{m}\big)^2} \nonumber \\
&&+\cos^2\big(|g|(\sqrt{n}+\sqrt{m})\tau\big) \ee^{-\frac{2|\gamma|^2}{\omega^2_t}
\Big(1-\cos(\omega_t \tau)\Big)\big(\sqrt{n}+\sqrt{m}\big)^2}+\Big(\cos\big(2|g|\sqrt{n}\tau\big) \nonumber \\
&&+\cos\big(2|g|\sqrt{m}\tau\big) \Big)\ee^{-\frac{2|\gamma|^2}{\omega^2_t}
\Big(1-\cos(\omega_t \tau)\Big)\big(n+m\big)} \Big] 
\ee^{-i\Big(\omega_c t-
\frac{2|\gamma|^2}{\omega^2_t}\left(\omega_t \tau-\sin(\omega_t \tau)\right)\Big)(n-m)}, \nonumber 
\end{eqnarray}
\begin{eqnarray}
&& a_{12}(n,m)=\frac{1}{16} \ee^{-|\alpha|^2} \frac{\alpha^{n+1} \big(\alpha^* \big)^{m+1}}{\sqrt{(n+1)!}\sqrt{(m+1)!}}
\Big[\cos^2\big(|g|(\sqrt{n+1}-\sqrt{m+1})\tau\big) \times \nonumber \\
&&\times\ee^{-\frac{2|\gamma|^2}{\omega^2_t}
\Big(1-\cos(\omega_t \tau)\Big)\big(\sqrt{n+1}-\sqrt{m+1}\big)^2}\nonumber \\
&&-\cos^2\big(|g|(\sqrt{n+1}+\sqrt{m+1})\tau\big)
\ee^{-\frac{2|\gamma|^2}{\omega^2_t}
\Big(1-\cos(\omega_t \tau)\Big)\big(\sqrt{n+1}+\sqrt{m+1}\big)^2} \Big] \times \nonumber \\ 
&& \times \ee^{-i\Big(\omega_c t-
\frac{2|\gamma|^2}{\omega^2_t}\left(\omega_t \tau-\sin(\omega_t \tau)\right)\Big)(n-m)}, \nonumber 
\end{eqnarray}
\begin{eqnarray}
&&a_{21}(n,m)=\frac{1}{16} \ee^{-|\alpha|^2} \frac{\alpha^{n+1} \big(\alpha^* \big)^{m+1}}{\sqrt{(n+1)!}\sqrt{(m+1)!}}
\Big[-\cos\big(|g|(\sqrt{n+1}+\sqrt{m+1})\tau\big)\times \nonumber \\ 
&&\cos\big(|g|(\sqrt{n}+\sqrt{m})\tau\big)
\ee^{-\frac{|\gamma|^2}{\omega^2_t}
\Big(1-\cos(\omega_t \tau)\Big)\{\big(\sqrt{n+1}+\sqrt{m+1}\big)^2+\big(\sqrt{n}+\sqrt{m}\big)^2\}} \nonumber \\
&&-\cos\big(|g|(\sqrt{n+1}+\sqrt{m+1})\tau\big)\cos\big(|g|(\sqrt{n}-\sqrt{m})\tau\big)\times \nonumber \\
&&\times\ee^{-\frac{|\gamma|^2}{\omega^2_t}
\Big(1-\cos(\omega_t \tau)\Big)\{\big(\sqrt{n+1}+\sqrt{m+1}\big)^2+\big(\sqrt{n}-\sqrt{m}\big)^2\}} \nonumber \\
&&+\cos\big(|g|(\sqrt{n+1}-\sqrt{m+1})\tau\big)\cos\big(|g|(\sqrt{n}+\sqrt{m})\tau\big)\times \nonumber \\
&&\times\ee^{-\frac{|\gamma|^2}{\omega^2_t}
\Big(1-\cos(\omega_t \tau)\Big)\{\big(\sqrt{n+1}-\sqrt{m+1}\big)^2+\big(\sqrt{n}+\sqrt{m}\big)^2\}} \nonumber \\
&&+\cos\big(|g|(\sqrt{n+1}-\sqrt{m+1})\tau\big)\cos\big(|g|(\sqrt{n}-\sqrt{m})\tau\big)\times \nonumber \\
&&\times\ee^{-\frac{|\gamma|^2}{\omega^2_t}
\Big(1-\cos(\omega_t \tau)\Big)\{\big(\sqrt{n+1}-\sqrt{m+1}\big)^2+\big(\sqrt{n}-\sqrt{m}\big)^2\}}
\Big] 
\ee^{-i\Big(\omega_c t-
\frac{2|\gamma|^2}{\omega^2_t}\left(\omega_t \tau-\sin(\omega_t \tau)\right)\Big)(n-m)}, \nonumber 
\end{eqnarray}
\begin{eqnarray} 
&&a_{22}(n,m)=\frac{1}{16} \ee^{-|\alpha|^2} \frac{\alpha^{n+2} \big(\alpha^* \big)^{m+2}}{\sqrt{(n+2)!}\sqrt{(m+2)!}}
\Big[\cos\big(|g|(\sqrt{n+2}+\sqrt{m+2})\tau\big) \nonumber \\
&&\cos\big(|g|(\sqrt{n+1}+\sqrt{m+1})\tau\big) \ee^{-\frac{|\gamma|^2}{\omega^2_t}
\Big(1-\cos(\omega_t \tau)\Big)\{\big(\sqrt{n+2}+\sqrt{m+2}\big)^2+\big(\sqrt{n+1}+\sqrt{m+1}\big)^2\}} \nonumber \\
&&-\cos\big(|g|(\sqrt{n+2}+\sqrt{m+2})\tau\big)\cos\big(|g|(\sqrt{n+1}-\sqrt{m+1})\tau\big) \times \nonumber \\
&&\times\ee^{-\frac{|\gamma|^2}{\omega^2_t}
\Big(1-\cos(\omega_t \tau)\Big)\{\big(\sqrt{n+2}+\sqrt{m+2}\big)^2+\big(\sqrt{n+1}-\sqrt{m+1}\big)^2\}} \nonumber \\
&&-\cos\big(|g|(\sqrt{n+2}-\sqrt{m+2})\tau\big)\cos\big(|g|(\sqrt{n+1}+\sqrt{m+1})\tau\big) \times \nonumber \\
&&\times\ee^{-\frac{|\gamma|^2}{\omega^2_t}
\Big(1-\cos(\omega_t \tau)\Big)\{\big(\sqrt{n+2}-\sqrt{m+2}\big)^2+\big(\sqrt{n+1}+\sqrt{m+1}\big)^2\}} \nonumber \\
&&+\cos\big(|g|(\sqrt{n+2}-\sqrt{m+2})\tau\big)\cos\big(|g|(\sqrt{n+1}-\sqrt{m+1})\tau\big)\times \nonumber \\
&&\times\ee^{-\frac{|\gamma|^2}{\omega^2_t}
\Big(1-\cos(\omega_t \tau)\Big)\{\big(\sqrt{n+2}-\sqrt{m+2}\big)^2+\big(\sqrt{n+1}-\sqrt{m+1}\big)^2\}}
\Big] 
\ee^{-i\Big(\omega_c t-
\frac{2|\gamma|^2}{\omega^2_t}\left(\omega_t \tau-\sin(\omega_t \tau)\right)\Big)(n-m)}. \nonumber
\end{eqnarray}

\end{document}